\documentclass[12pt]{article}
\usepackage{epsfig,cite}
\include{epsf}
\newcount\timecount
\newcount\hours \newcount\minutes  \newcount\temp \newcount\pmhours
\hours = \time
\divide\hours by 60
\temp = \hours
\multiply\temp by 60
\minutes = \time
\advance\minutes by -\temp
\def\hour{\the\hours}
\def\minute{\ifnum\minutes<10 0\the\minutes
            \else\the\minutes\fi}
\def\clock{
\ifnum\hours=0 12:\minute\ AM
\else\ifnum\hours<12 \hour:\minute\ AM
      \else\ifnum\hours=12 12:\minute\ PM
            \else\ifnum\hours>12
                 \pmhours=\hours
                 \advance\pmhours by -12
                 \the\pmhours:\minute\ PM
                 \fi
            \fi
      \fi
\fi
}

\def\monthname{\relax\ifcase\month 0/\or January\or February\or
   March\or April\or May\or June\or July\or August\or September\or
   October\or November\or December\else\number\month/\fi}

\def\bold#1{\setbox0=\hbox{$#1$}%
     \kern-.025em\copy0\kern-\wd0
     \kern.05em\copy0\kern-\wd0
     \kern-.025em\raise.0433em\box0 }

\def\beq{\begin{equation}}
\def\eeq{\end{equation}}

\def\ga{\mathrel{\raise.3ex\hbox{$>$\kern-.75em\lower1ex\hbox{$\sim$}}}}
\def\la{\mathrel{\raise.3ex\hbox{$<$\kern-.75em\lower1ex\hbox{$\sim$}}}}
\def\gev{{\rm \, Ge\kern-0.125em V}}
\def\tev{{\rm \, Te\kern-0.125em V}}
\def\gyr{{\rm \, G\kern-0.125em yr}}

\def\gappeq{\mathrel{\rlap {\raise.5ex\hbox{$>$}}
{\lower.5ex\hbox{$\sim$}}}}
\def\lappeq{\mathrel{\rlap{\raise.5ex\hbox{$<$}}
{\lower.5ex\hbox{$\sim$}}}}
\def\Toprel#1\over#2{\mathrel{\mathop{#2}\limits^{#1}}}

\def\m12{m_{1\!/2}}

\def\PL{{Phys.Lett.} }
\def\PR{{Phys.Rev.} }

\def\bea{\begin{eqnarray}}
\def\eea{\end{eqnarray}}

\begin{document}
\begin{titlepage}
\pagestyle{empty}
\baselineskip=21pt
\rightline{\tt hep-ph/0408118}
\rightline{CERN-PH-TH/2004-131}
\rightline{UMN--TH--2315/04}
\rightline{FTPI--MINN--04/27}
\vskip 0.2in
\begin{center}
{\large {\bf Prospects for Sparticle Discovery in Variants of the MSSM}}
\end{center}
\begin{center}
\vskip 0.2in
{\bf John~Ellis}$^1$, {\bf Keith~A.~Olive}$^{2}$, {\bf Yudi~Santoso}$^{2}$ 
and {\bf Vassilis~C.~Spanos}$^{2}$
\vskip 0.1in
{\it
$^1${TH Division, CERN, Geneva, Switzerland}\\
$^2${William I. Fine Theoretical Physics Institute, \\
University of Minnesota, Minneapolis, MN 55455, USA}}\\
\vskip 0.2in
{\bf Abstract}
\end{center}
\baselineskip=18pt \noindent

We discuss the prospects for detecting supersymmetric particles in
variants of the minimal supersymmetric extension of the Standard Model
(MSSM), in light of laboratory and cosmological constraints. We first
assume that the lightest supersymmetric particle (LSP) is the lightest
neutralino $\chi$, and present scatter plots of the masses of the two
lightest visible supersymmetric particles when the input scalar and
gaugino masses are constrained to be universal (CMSSM), when the input
Higgs scalar masses are non-universal (NUHM), and when the squark and
slepton masses are also non-universal and the MSSM is regarded as a low-energy
effective field theory valid up to the GUT scale (LEEST) or just up to
10~TeV (LEEST10). We then present similar plots in various scenarios when
the LSP is the gravitino. We compare the prospects for detecting
supersymmetry at linear colliders (LCs) of various energies, at the LHC,  and as
astrophysical dark matter. We find that, whilst a LC with a centre-of-mass
energy $E_{CM} \le 1000$~GeV has some chance of discovering the lightest
and next-to-lightest visible supersymmetric particles, $E_{CM} \ge 3000$~GeV would be
required to `guarantee' finding supersymmetry in the neutralino LSP
scenarios studied, and an even higher $E_{CM}$ might be required in
certain gravitino dark matter scenarios.  Direct dark matter experiments
could explore part of the low-mass neutralino LSP region, but would not
reveal all the models accessible to a low-energy LC.

\vfill
\leftline{CERN-PH-TH/2004-131}
\leftline{August 2004}
\end{titlepage}
\baselineskip=18pt

\section{Introduction}

When considering projects for new high-energy accelerators, the prospects
for discovering supersymmetry are among the issues frequently considered.  
Since even the minimal supersymmetric extension of the Standard Model
(MSSM) has over 100 free parameters including those characterizing
supersymmetry breaking, these prospects are difficult to assess globally
in a convincing way, and simplifying assumptions are often made. A common
assumption is that $R$ parity is conserved, in which case the lightest
supersymmetric particle (LSP) is stable, and a possible candidate for the
cold dark matter postulated by astrophysicists and
cosmologists~\cite{EHNOS}. The LSP presumably has no strong or
electromagnetic interactions, but otherwise its nature is ambiguous. It is
often assumed that the LSP is the lightest neutralino $\chi$, but another
generic possibility is that the LSP is the gravitino ${\tilde
G}$~\cite{ekn} - \cite{feng}.

We consider both possibilities in this paper, constraining them using
laboratory, astrophysical and cosmological data. Specifically, we require
that the constraints from colliders (particularly LEP) and $b \to s
\gamma$ be obeyed~\footnote{Note that we do not apply any constraint from
$g_\mu - 2$, though we comment below on the possible effect of this
constraint.}, as well as the constraints from WMAP and other cosmological
data on the cold dark matter density, and (in the case of a gravitino LSP)
we require consistency between the baryon-to-entropy ratio inferred from
Big-Bang nucleosynthesis (BBN) and the cosmic microwave background 
(CMB)~\cite{CEFO}.

The impacts of these constraints are often explored in the framework of
the CMSSM, in which the input scalar and gaugino masses are constrained to
be universal, and the LSP is assumed to be the lightest
neutralino~\cite{us,them,djou}. We also include this scenario in our analysis,
but our scope is broader, since we also analyze neutralino LSP models in
which the input Higgs scalar masses are allowed to be non-universal (NUHM)
\cite{nonu,nuhm}, and in which the squark and slepton masses are also
non-universal and the MSSM is regarded as a low-energy effective theory
(LEEST) \cite{LEEST}.  We also consider gravitino dark matter models
(GDMs) in which different assumptions are made about the gravitino mass
relative to the input scalar and gaugino masses \cite{BBB,gdm,feng}.

In each case, we make a scatter plot of the masses of the lightest visible
supersymmetric particle (LVSP) and the next-to-lightest visible
supersymmetric particle (NLVSP). We do not consider the LSP itself to be
visible, nor any heavier neutral sparticle that decays invisibly inside
the detector, such as ${\tilde \nu} \to \nu \chi$  when ${\tilde \nu}$ is the 
next-to-lightest sparticle in a 
neutralino LSP scenario~\footnote{However, when the sneutrino has visible
decays it is regarded as a possible NLVSP.}, or is metastable and
decays outside the detector, such as $\chi \to \gamma {\tilde G}$ in a GDM
scenario. The LVSP and the NLVSP are the lightest sparticles likely to be
observable in collider experiments. Since the masses of the selectron and 
smuon are identical in all the (simplified) models we study, one would 
actually get `two for the price of one' in cases where a charged slepton 
is the LVSP or NLVSP.

At a generic linear $e^+ e^-$ collider (LC), the physics reach for any
visible supersymmetric particle is likely to be a mass close to the beam
energy. As is apparent from the scatter plots shown later in this paper, a
LC with $E_{CM} = 500$~GeV has some chance of producing and detecting one
or two sparticle types, particularly in models obeying the cosmological
and astrophysical constraints, but this cannot be guaranteed. A LC with
$E_{CM} = 1000$~GeV clearly has a greater chance of producing sparticles,
but this still cannot be guaranteed. Only a LC with $E_{CM} = 3000$~GeV
seems `guaranteed' to produce and detect sparticles, within the variants
of the MSSM with a neutralino LSP studied here, namely the CMSSM, NUHM,
LEEST and LEEST10, but an even higher $E_{CM}$ might be required in some
GDM scenarios. For related studies, see \cite{related}.

For comparison, we also indicate the range of neutralino LSP models in
which supersymmetric dark matter may be observable directly in elastic
scattering experiments, assuming a sensitivity to the spin-independent
$\chi-$N scattering cross section $\ga 10^{-8}$~pb. We find that some
fraction of the models with a light neutralino LSP that are accessible to
a low-energy LC might give an observable dark matter signal, but not all.
Thus, a low-energy LC would add value by exploring the low-mass part of
the parameter space more completely.

\section{Methodology}

Our procedure for analyzing the parameter spaces in each of the
supersymmetric models we study is to generate a sample with 50,000 random
choices of mass parameters, up to an upper limit of 2~TeV for the soft
supersymmetry-breaking squark and slepton mass parameters $m_Q, m_D, m_U$
and $m_L, m_E$. We also allow the gaugino mass parameter $m_{1/2}$ (which
is assumed to be universal for the SU(3), SU(2) and U(1) factors) to vary over
this range. The soft Higgs masses $m_{1,2}^2$ are varied from -4 to 4 TeV$^2$. 
The physical
Higgs masses squared, which include both the soft supersymmetry-breaking
contribution and the $\mu$-dependent contribution, are constrained to be
positive up to some high energy scale (either the GUT scale or 10 TeV as
described below). We allow the trilinear soft supersymmetry breaking
parameter $A_0$ to vary over the range $-1~{\rm TeV} < A_0 < 1$~TeV. We
treat $|\mu|$ and the pseudoscalar Higgs mass $m_A$ as dependent
parameters that are fixed by the electroweak vacuum conditions. The
arbitrary upper limits on the mass parameters are crude reflections of the
upper limits that are supposed to be motivated by naturalness
arguments~\cite{EENZ}. However, in many of the models under study, they
are ample to include all the models that obey the cosmological constraints
described below. We sample $1.8 < \tan \beta < 58$ for $\mu > 0$ and $1.8
< \tan \beta < 43$ for $\mu < 0$:  above these upper limits, we no longer
find solutions of the electroweak vacuum conditions in generic regions of
parameter space.

Our procedure for implementing the laboratory constraints on
supersymmetric models follows that described elsewhere \cite{us}. The most relevant
constraints are those due to the LEP lower limits on the chargino mass
$m_{\chi^\pm}$ and the Higgs mass $m_h$, and the agreement of $b \to s
\gamma$ decay with the prediction of the Standard Model, within
experimental and theoretical errors. Note that we use here the recent
update on the top-quark mass \cite{mtop}, $m_t = 178$~GeV, which has a significant
impact on the interpretations of the Higgs limit in the various model
parameter spaces. For example, in the CMSSM, the increase from $m_t =
175$~GeV decreases the lower limit on the universal gaugino mass from
$m_{1/2} \sim 300$~GeV to $\sim 250$~GeV for $\tan \beta = 10$ as calculated
using FeynHiggs \cite{FeynHiggs}. Changing
$m_t$ has other important impacts on model parameter spaces, such as
moving rapid-annihilation poles \cite{funnels} and focus-point regions \cite{fp}. 
While the former are certainly included in our samples, the sensitivity of 
the focus point is well known~\cite{rs}, and it is
pushed to values of $m_0$ far beyond 
our sampling range. For example, at $m_{1/2} = 300, \tan \beta = 10$, and 
$A_0 = 0$, we find that the focus point moves from $\sim 2.5$~TeV to 
greater than 4.8~TeV when $m_t$ is increased from 175~GeV to 178~GeV. 
Bearing in mind this sensitivity of the focus-point region and the fact 
that it lies beyond our sampling range for our default choice of $m_t$, we 
do not discuss it further in this paper.  We do note however, that unless
our range for $m_{1/2}$ is increased, the focus point would yield
a LVSP and NLVSP which is either a neutralino or chargino
and would not go beyond the bounds already considered.

We do not take explicitly into account the possible constraint from $g_\mu
- 2$~\cite{g-2}, in view of the persistent uncertainties in the estimate 
of the
contribution from hadronic vacuum polarization. However, we do note that,
generically, the regions of the parameter spaces with $\mu > 0$ are
normally compatible with experiment at the 2-$\sigma$ level. Including
this constraint would have very little effect the models we display for
$\mu >0$, and the constraint would have no effect at all on models with
large LVSP and NLVSP masses.  In contrast, regions with $\mu < 0$ are
normally incompatible with $g_\mu - 2$ at the 2-$\sigma$ level, and
essentially all models shown for $\mu < 0$ are excluded by the $g_\mu-2$
constraint. Thus, although we show results for both signs of $\mu$, only
positive values of $\mu$ are formally consistent with this constraint.

Our procedures for implementing cosmological and astrophysical constraints
also follow those discussed elsewhere \cite{us}. For the cold dark matter
density, we use the range $0.094 < \Omega_{CDM} h^2 < 0.129$ preferred by
a joint analysis of first-year WMAP and other data \cite{wmap}. In the
case of neutralino LSP models, we identify $\Omega_{CDM} = \Omega_\chi$:
allowing other contributions to $\Omega_{CDM}$~\footnote{These might arise
from non-thermal mechanisms such as moduli decays in specific scenarios
for supersymmetric cosmology \cite{kmy}.} would, in general, allow also somewhat
smaller sparticle masses, but the effect is not large. In the case of GDM,
we require the density of gravitinos produced in the decays of heavier
sparticles not to exceed the upper limit $\Omega_{CDM} h^2 = 0.129$, but
we do allow values below 0.094, since gravitinos are likely to have also
been produced by generic thermal or other mechanisms in the very early
Universe. A further important constraint on GDM scenarios is that on the
Standard Model decay products $X$ accompanying the decays of sparticles
${\tilde Y}$ into gravitinos: ${\tilde Y} \to X + {\tilde G}$. These
cannot perturb greatly the abundances of light elements, since
astrophysical observations agree with their abundances calculated from
Big-Bang nucleosynthesis using the baryon-to-entropy ratio inferred from
WMAP and other measurements of the cosmic microwave background (CMB). We
implement this constraint following the analysis in~\cite{gdm,CEFO}.

We close this section with some comments on the possible natures of the
LVSP and NLVSP. In different regions of the parameter spaces for
neutralino LSP models these might include the lighter stau ${\tilde
\tau_1}$, the $({\tilde e}_R, {\tilde \mu}_R)$, the lightest chargino
$\chi^\pm$ or the second neutralino $\chi_2$, and in GDM models the
lightest neutralino $\chi$ also becomes a candidate. Depending on the
model, the ${\tilde \tau_1}$ may have quite a different mass from the
${\tilde e}_R$ and ${\tilde \mu}_R$, but the latter are 
degenerate in our analysis, because we assume degenerate sfermion 
masses before renormalization and neglect the $e$ and $\mu$ Yukawa 
couplings. Thus, in parameter regions where these are the LVSP or NLVSP, 
one actually observes two sparticles for the price of one.

\section{Results for Collider Searches}

\begin{figure}
\begin{center}
\mbox{\epsfig{file=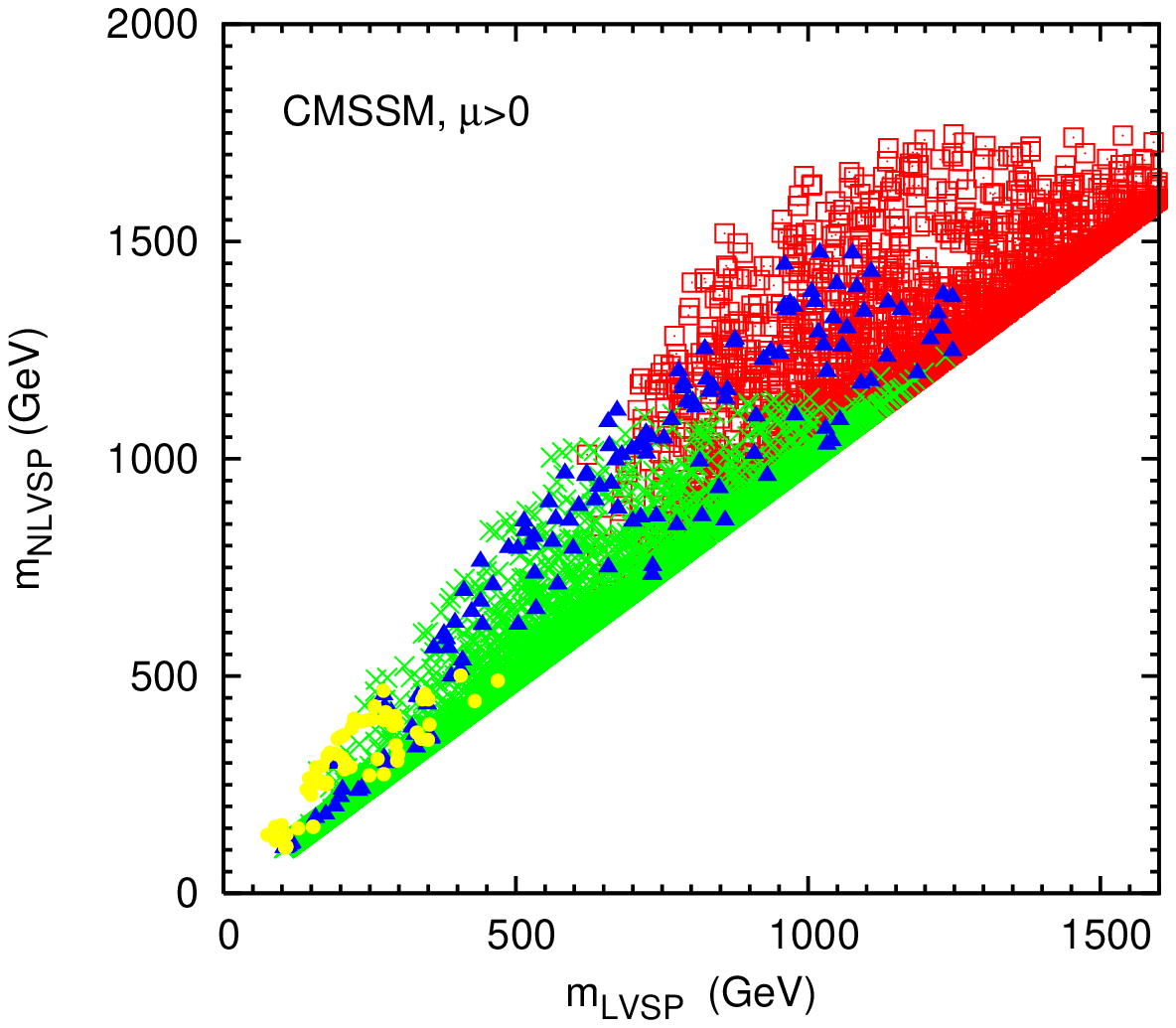,height=6.8cm}}
\mbox{\epsfig{file=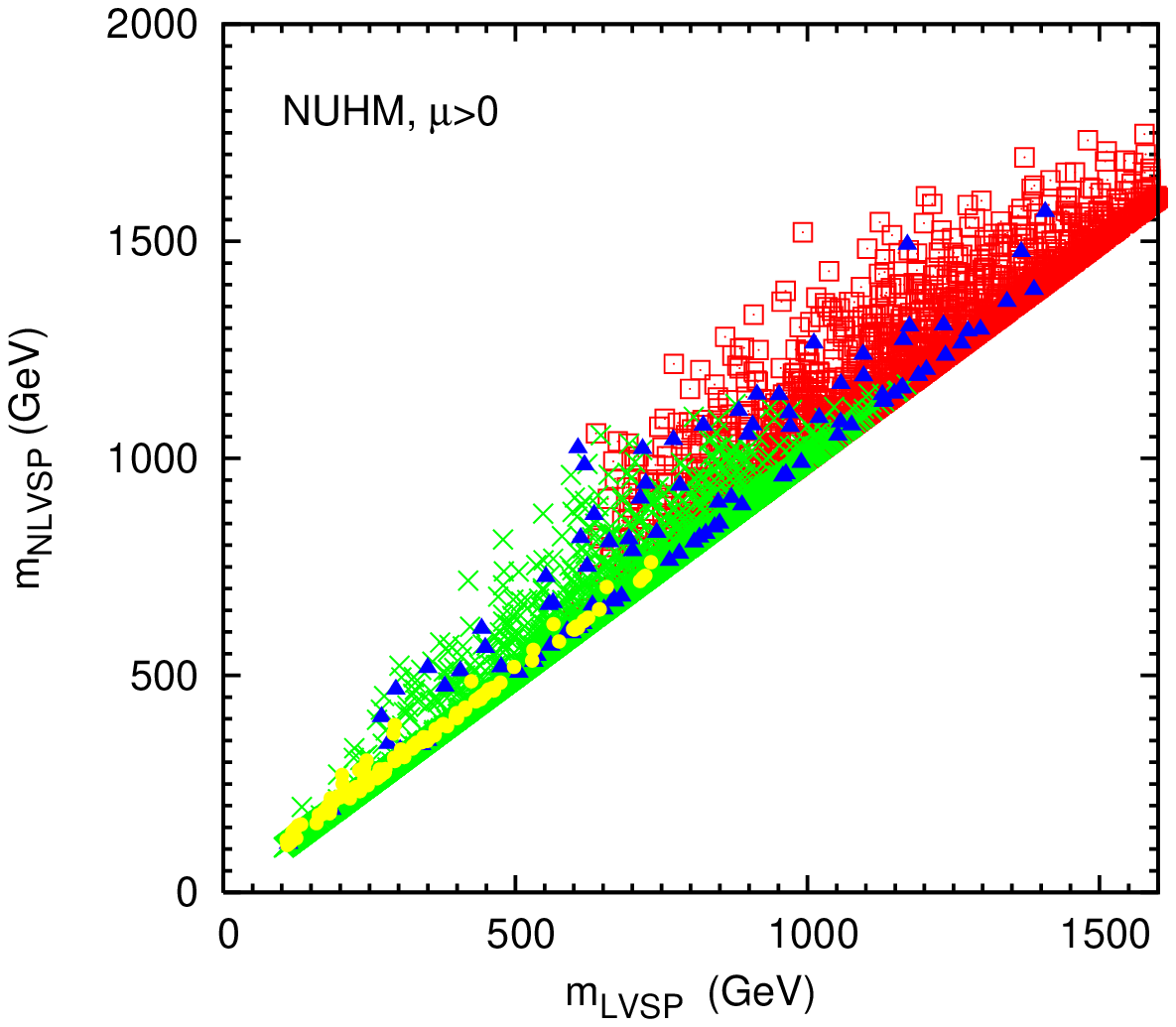,height=6.8cm}}
\end{center} 
\begin{center}
\mbox{\epsfig{file=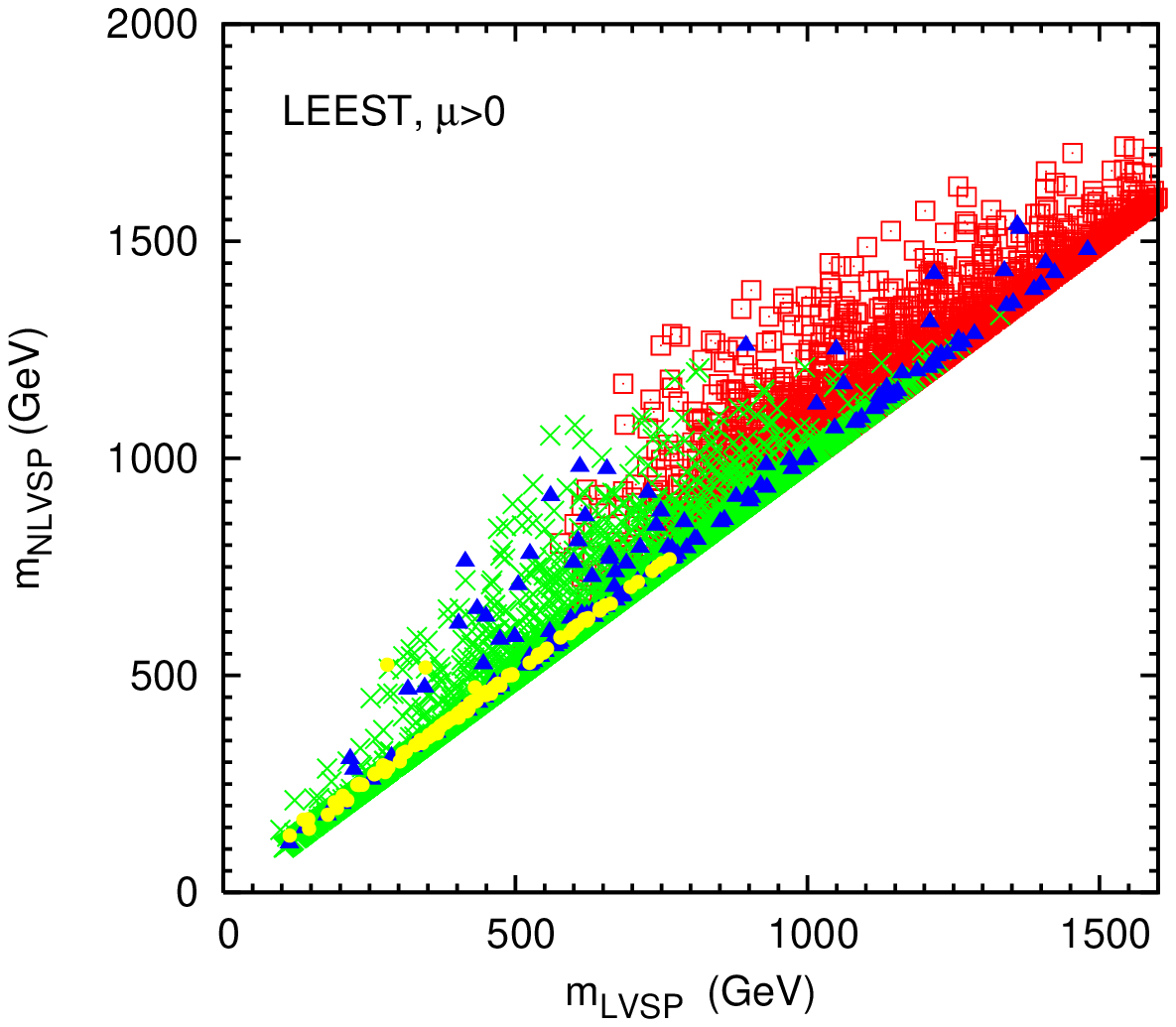,height=6.8cm}}
\mbox{\epsfig{file=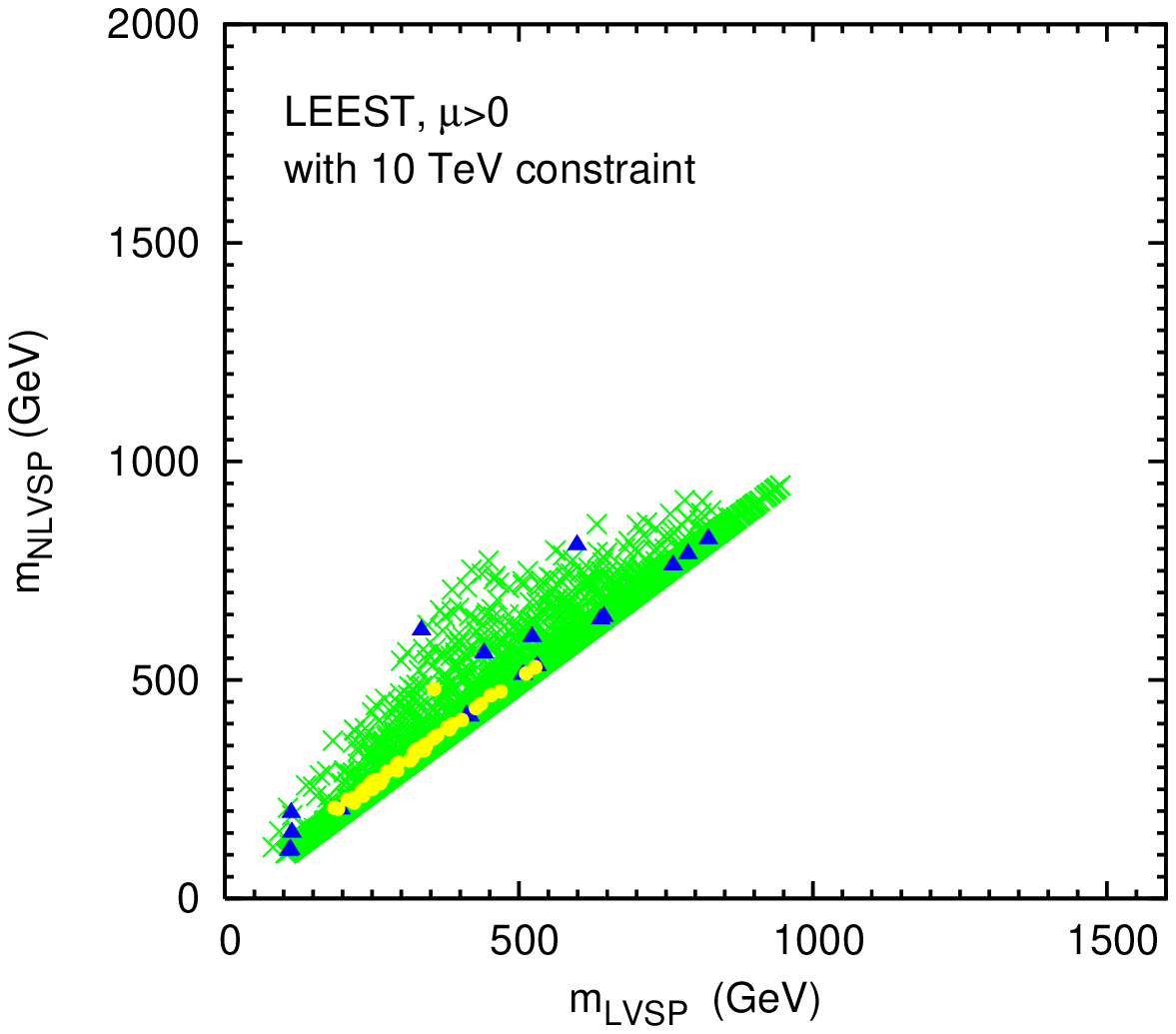,height=6.8cm}}
\end{center} 
\caption{\it
Scatter plots of the masses of the lightest visible
supersymmetric particle (LVSP) and the next-to-lightest visible
supersymmetric particle (NLVSP) in (a) the CMSSM, (b) the NUHM, (c) the 
LEEST and (d) the LEEST10, all for $\mu > 0$. The darker (blue) 
triangles satisfy all the laboratory, astrophysical and cosmological 
constraints. For comparison, the dark (red) squares and medium-shaded 
(green) crosses respect the laboratory 
constraints, but not those imposed by astrophysics and cosmology. 
In addition, the (green) crosses represent models which are expected to be
visible at the LHC. The 
very light (yellow) points are those for which direct detection of 
supersymmetric dark matter might be possible according to the criterion 
discussed in the text.} 
\label{fig:VSP10p} 
\end{figure}

Our first set of results is shown in Fig.~\ref{fig:VSP10p} for the choice
$\mu > 0$, with panel (a) displaying our findings for the CMSSM.  All
points shown satisfy the phenomenological constraints discussed above.  
The dark (red) squares represent those points for which the relic density
is outside the WMAP range, and for which all coloured sparticles (squarks
and gluinos) are heavier than 2 TeV. The CMSSM parameter reach at the LHC
has been analyzed in~\cite{Baer}, which used ISAJET~v7.64 and
CMSJET~v4.801 to simulate the prospective CMS signals in many channels. To
within a few percent accuracy, the CMSSM reach contours presented
in~\cite{Baer} for different choices of $\tan \beta$ and the sign of $\mu$
coincide with the 2-TeV contour for the lightest squark (generally the
stop) or gluino, so we regard the dark (red) points as unobservable at the
LHC. Most of these points have $m_{NLVSP} \ga 1.2$~TeV. Conversely, the
medium-shaded (green) crosses represent points where at least one squark
or gluino has a mass less than 2 TeV and should be observable at the LHC,
according to~\cite{Baer}. The spread of the dark (red) squares and
medium-shaded (green) crosses, by as much as 500~GeV or more in some
cases, reflects the maximum mass splitting between the LVSP and the NLVSP
that is induced in the CMSSM via renormalization effects on the input mass
parameters. The amount of this spread also reflects our cutoff $|A_0| < 1$
TeV, which controls the mass splitting of the third generation sfermions.
 
The darker (blue) triangles are those points respecting the cosmological
cold dark matter constraint~\footnote{We see in the bottom-left part of
this and subsequent scatter plots some lighter (yellow) points which also have
$\Omega_{CDM} h^2 < 0.129$, but may have
$\Omega_{CDM} h^2 < 0.094$.}. Comparing with the regions populated by dark
(red)  squares and medium-shaded (green) crosses, one can see which of
these models would be detectable at the LHC, according to the criterion in
the previous paragraph. We see immediately that the dark matter constraint
restricts the LVSP masses to be less than about 1250~GeV and NLVSP masses
to be less than about 1500~GeV. In most cases, the identity of the LVSP
is the lighter $\tilde \tau$. While pair-production of the LVSP would
sometimes require a CM energy of about 2.5~TeV, in some cases there 
is a lower supersymmetric threshold due to
the associated production of the LSP $\chi$ with the 
next lightest 
neutralino $\chi_2$~\cite{djou}. Examining the masses and 
identities of the sparticle
spectrum at these points, we find that $E_{CM} \ga 2.2$ TeV would be 
sufficient to see 
at least one sparticle, as shown in Table~1. Similarly,  only a LC with 
$E_{CM} \ge
2.5$~TeV would be `guaranteed' to see two visible sparticles (in addition
to the $\chi$ LSP), somewhat lower than the 3.0 TeV one might
obtain by requiring the pair production of the NLVSP. We note that, in this 
and other cases, it is 
possible that some points with higher $m_{LVSP}$ and/or $m_{NVSP}$ might 
be found in a larger 
sample of models. Larger masses may occur in the focus-point region, as 
noted 
above, as well as when the neutralino and some other 
sparticle  are nearly degenerate (such as the stop when $A$ is large)
and coannihilation controls the relic LSP 
density~\cite{stopco}~\footnote{This is 
just one reason why our `guarantees' are in quotation 
marks.}. Our points with $m_{LVSP} \ga 700$ GeV are
predominantly due to rapid annihilation via direct-channel $H,A$ poles,
while points with 200 GeV $\la m_{LVSP} \la$ 700 GeV are largely due to
$\chi$-slepton coannihilation. If either of these effects were
overlooked, the upper limits on $m_{LVSP}$ and $m_{NLVSP}$ would be
considerably tighter.

\begin{table}[htb]
\begin{center}
\caption{\it Centre-of-mass energy (in TeV) required to observe 
one or two sparticles at a future LC in each of the models discussed in 
the 
text.}
\label{tab:alpha}
\vskip .3cm
\begin{tabular}{|c|c|c|c|}
\hline {\it Model} &
$sgn(\mu)$  & {\it one sparticle} & {\it two sparticles} \\
\hline CMSSM & $\mu > 0 $ &  2.2  & 2.6  \\
&$\mu < 0$& 2.2 & 2.5 \\ \hline
 NUHM & $\mu > 0 $ &  2.4  & 2.8 \\
&$\mu < 0$& 2.6& 2.9 \\ \hline
 LEEST & $\mu > 0 $ & 2.6  & 3.0 \\
&$\mu < 0$& 2.5& 3.2 \\  \hline
 LEEST10 & $\mu > 0 $ &  1.2  &1.6 \\
&$\mu < 0$& 1.1& 1.5 \\  \hline
 GDM $m_{3/2} = 10$ GeV & $\mu > 0 $ &  1.1  & 1.7  \\
&$\mu < 0$& 1.1& 1.4 \\ \hline
  GDM $m_{3/2} = 100$ GeV & $\mu > 0 $ &  2.6  & 2.9 \\
&$\mu < 0$& 2.6& 3.5 \\ \hline
  GDM $m_{3/2} = 0.2 m_0$ & $\mu > 0 $ & 2.5  &2.7 \\
&$\mu < 0$& 2.6& 3.0 \\  \hline
 GDM $m_{3/2} = m_0$  & $\mu > 0 $ &  1.7  &1.8 \\
&$\mu < 0$& 1.7 & 1.9 \\  \hline
\end{tabular}
\end{center}
\end{table}

An $E_{CM} = 500$~GeV LC would be able to explore the `bulk' region at low
$(m_{1/2}, m_0) $, which is represented by the small cluster of points
around $m_{LVSP} \sim 200$ GeV. It should also be noted that there are a
few points with $m_{LVSP} \sim 100$ GeV which are due to rapid
annihilation via the light Higgs pole. These points all have very large
values of $m_0$ which relaxes the Higgs mass and chargino mass
constraints, particularly when $m_t = 178$ GeV. A LC with $E_{CM} =
1000$~GeV would be able to reach some way into the coannihilation `tail',
but would not cover all the WMAP-compatible dark (blue) triangles. Indeed,
about a third of these points are even beyond the reach of the LHC in this
model. Finally, the light (yellow) filled circles are points for which the
elastic $\chi$-$p$ scattering cross section is larger than $10^{-8}$~pb.
All of these points have $\Omega h^2 < 0.129$. For those points with
$\Omega h^2 < 0.0945$, the cross section has been scaled downward by
$\Omega h^2/.0945$, to allow for another component of cold dark matter
which populates proportionally our galactic halo. We discuss these points
in more detail in the next section.

Panel (b) of Fig.~\ref{fig:VSP10p} displays a corresponding scatter plot
for the NUHM, in which the soft supersymmetry-breaking masses of the Higgs
bosons are allowed to float relative to those of the squarks and sleptons,
which are still assumed to be universal. We again use the 2-TeV mass
criterion motivated by~\cite{Baer} to distinguish models that are
unobservable at the LHC (dark, red) from those that are unobservable. No
analysis as detailed as~\cite{Baer} has been made in the NUHM, but we do
not expect large differences from the CMSSM. The `footprint' of
the darker (blue) points that respect the cosmological cold dark matter
constraint is similar in shape and origin from that in the CMSSM shown in
panel (a).  Once again, the dark (blue) triangles with large masses are
predominantly due to rapid $s$-channel annihilation through the $H,A$
poles. Because we allow the two soft Higgs masses to take values different
from $m_0$, $\mu$ and $m_A$ take on a significantly broader range of
values in the NUHM as compared to the CMSSM.  Thus, the rapid annihilation
funnels appear more frequently at all values of $\tan \beta$, in contrast
to the CMSSM, where the funnels appear only at high $\tan \beta$.  The
nearly linear track of points with $m_{LVSP} \simeq m_{NLVSP}$ corresponds
to points with large $m_0$ for which the LVSP and NLVSP are a nearly
degenerate pair of charginos and neutralinos. Points with smaller $m_0$
are dispersed to higher $m_{NLVSP} $ where the LVSP, NLVSP pair is
typically the stau and the selectron/smuon.

The LVSP could be as heavy as $\sim$ 1400~GeV and the NLVSP as heavy as $\sim$ 1600~GeV
in the NUHM case.  In the NUHM, production of a $\chi_1, \chi_2$ pair at 
a LC with $E_{CM} \ge 2.4$ TeV is sufficient to guarantee 
the detection of  at least one visible sparticle (in addition to the
$\chi$ LSP), whilst only a LC with $E_{CM} \gappeq 
2.8$ TeV (corresponding to the pair production of the LVSP) would be
`guaranteed' to see at least two visible sparticles. As in panel (a), a LC
with $E_{CM} \sim 500$~GeV or 1000~GeV would see sparticles in only a
corner of the overall footprint, though this might be the portion favoured
by some naturalness arguments. Also as before, we note that a low-energy 
LC would be able to spot models inaccessible to direct searches for dark 
matter.

Panels (c,d) of Fig.~\ref{fig:VSP10p} display the corresponding scatter
plots for the LEEST, in which no universality is assumed between the soft
supersymmetry-breaking squark and slepton masses with different gauge
quantum numbers. On the other hand, as motivated but not mandated by upper
limits on flavour-changing neutral interactions~\cite{ENNC}, we do assume
universality between squarks and sleptons that have the same gauge quantum
numbers but are in different generations. We require that the low-energy
effective supersymmetric theory remain viable, with a stable electroweak
vacuum, all the way up to some higher energy scale, taken in panel (c) to
be the GUT scale (LEEST) and 10~TeV (LEEST10) in panel
(d)~\footnote{Compared with~\cite{LEEST}, one technical difference is that
here the random sample is generated with input parameters at the high
scale, which are then run down to low scales using the
renormalization-group equations, whereas previously the random sample was
generated at the electroweak scale. This does not affect the conclusions
in any essential way.}. While the identity of the LVSP, NLVSP pair is
predominantly a chargino and neutralino or a stau, selectron/smuon pair as
in the NUHM, many other combinations are possible now. For example, one of
the sneutrinos is often the NLVSP. For LEEST10, we only require the theory
to remain viable up to 10 TeV, and we have made the analogous restriction
that scalar masses (at 10 TeV) lie between 0 and 2 TeV.  This constraint
removes many of the points from the initial set of data. This is the
reason for the paucity of points in panel (d).  This constraint further
makes it highly likely that at least one coloured sparticle exists with a
mass below 2 TeV, thus making all points potentially observable at the
LHC.

The conclusions to be drawn from the LEEST panel (c) do not differ
qualitatively from those in the CMSSM and NUHM panels (a,b): we use the
same criterion~\cite{Baer} for observability at the LHC, and the upper
limits on the LVSP and the NLVSP are about 1500~GeV.
Including $\chi_1, \chi_2$ production, the LEEST parameter space
scanned here could be covered by a LC with $E_{CM} > 2.6$ TeV 
(one sparticle) and $E_{CM} > 3.0$ TeV 
(two sparticles), as seen in Table~1. On the other hand, both the
darker (blue)  and lighter (green) points in panel (d) for the LEEST10
model extend up to somewhat smaller masses than seen previously: $m_{LVSP}
\sim 850~{\rm GeV}, m_{NLVSP} \sim 850$~GeV. This is due to the fact that
the renormalization of the soft supersymmetry-breaking parameters between
10~TeV and the electroweak scale is considerably less than that between
the GUT scale and the electroweak scale. For this reason, sparticle masses
are generally larger in LEEST than in LEEST10. Correspondingly, one would
be more optimistic about the physics reach of lower-energy LC if one did
not require the MSSM to remain valid all the way up to the GUT scale.
In this case, a LC with $E_{CM} > 1.2$ TeV 
(one sparticle) and $E_{CM} > 1.6$ TeV 
(two sparticles) is sufficient.

\begin{figure}
\begin{center}
\mbox{\epsfig{file=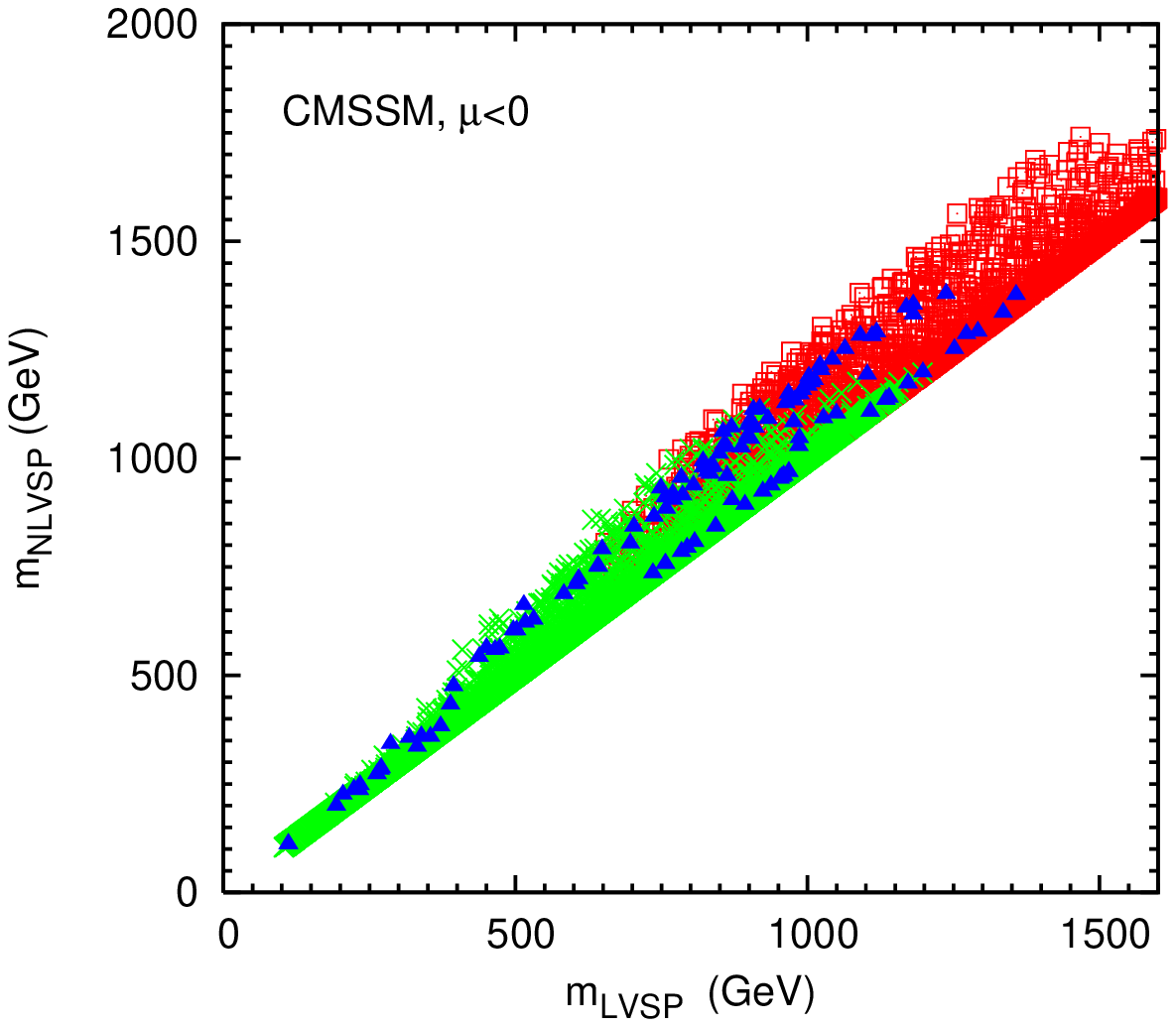,height=6.8cm}}
\mbox{\epsfig{file=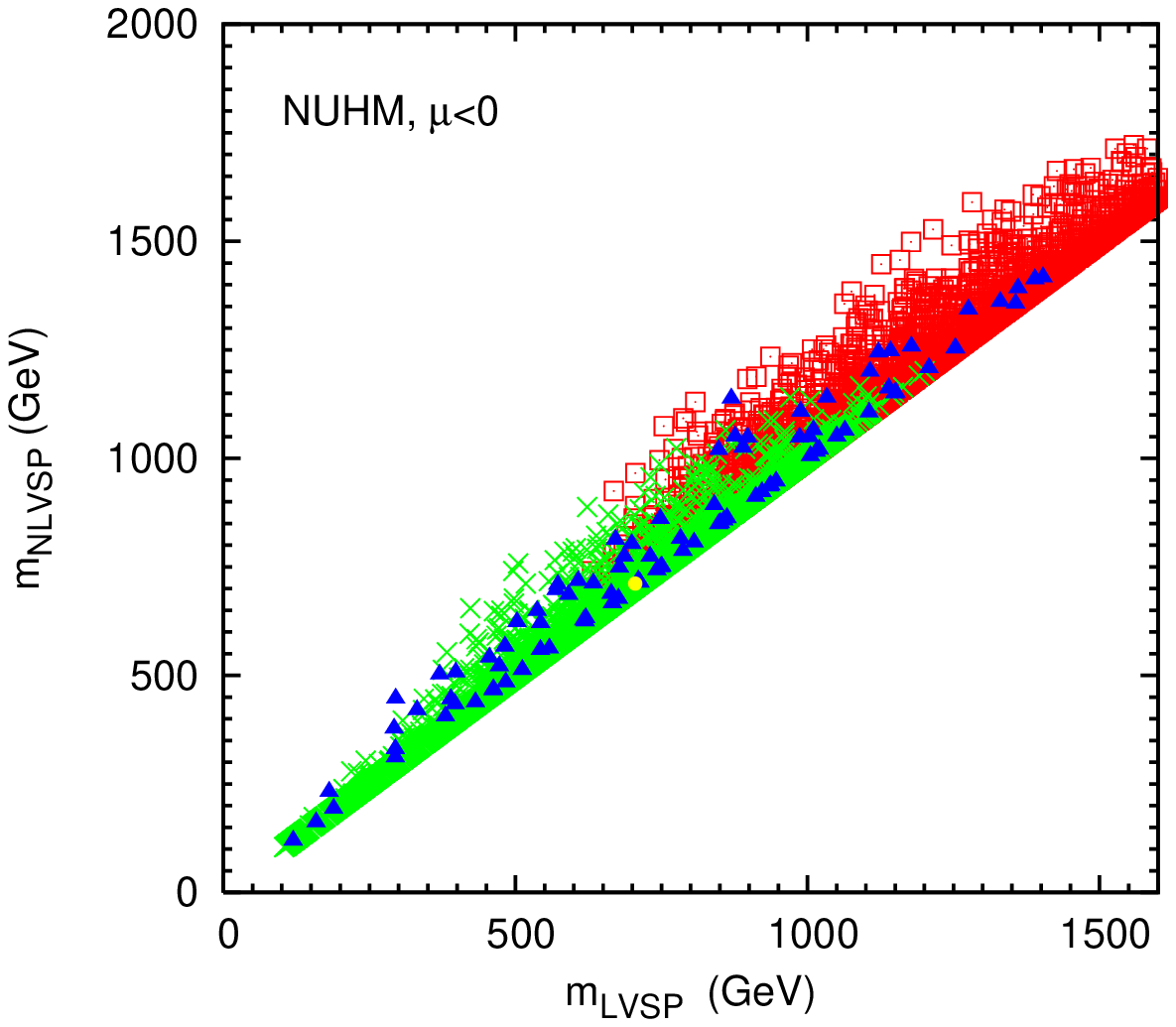,height=6.8cm}}
\end{center}
\begin{center}
\mbox{\epsfig{file=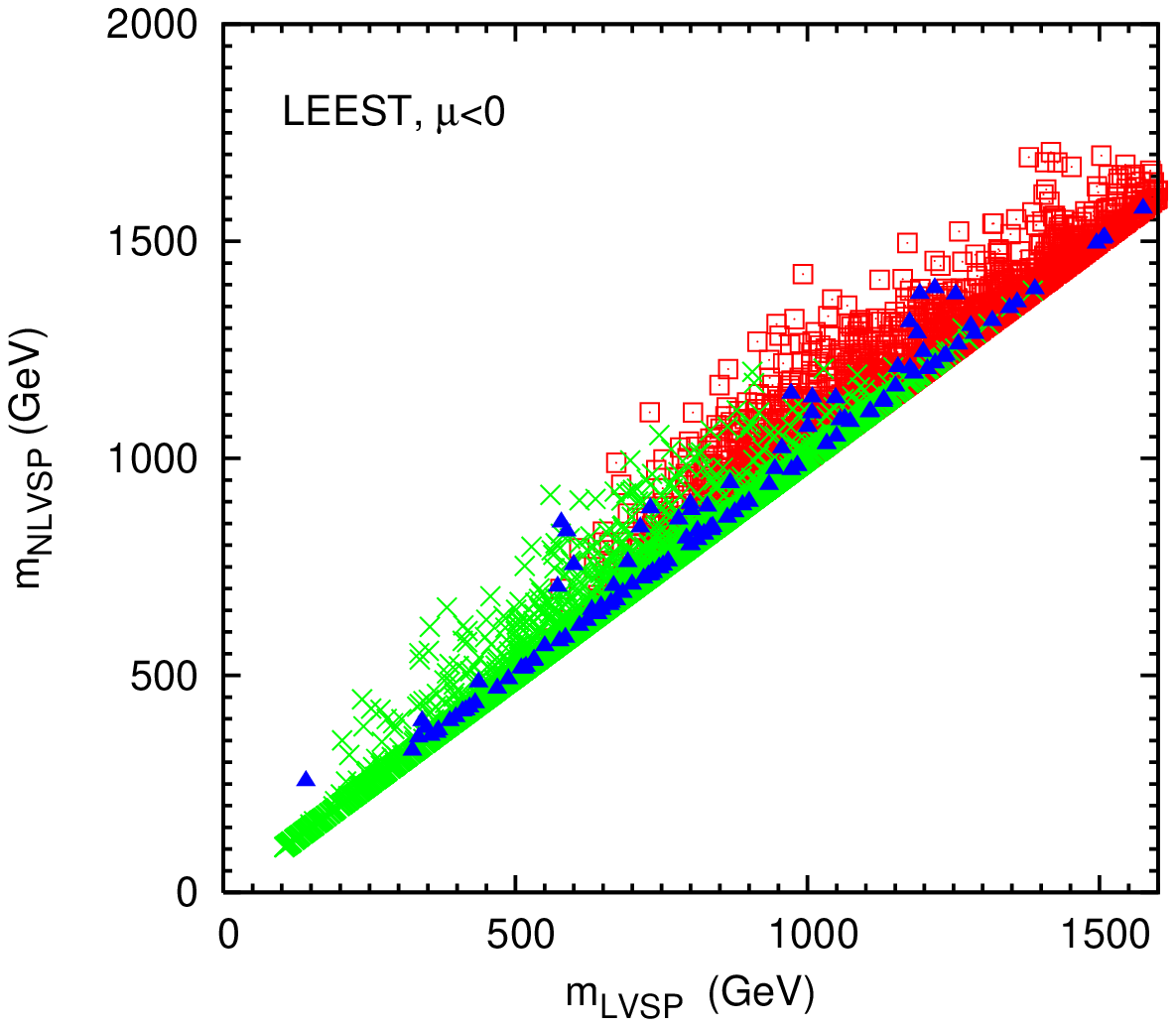,height=6.8cm}}
\mbox{\epsfig{file=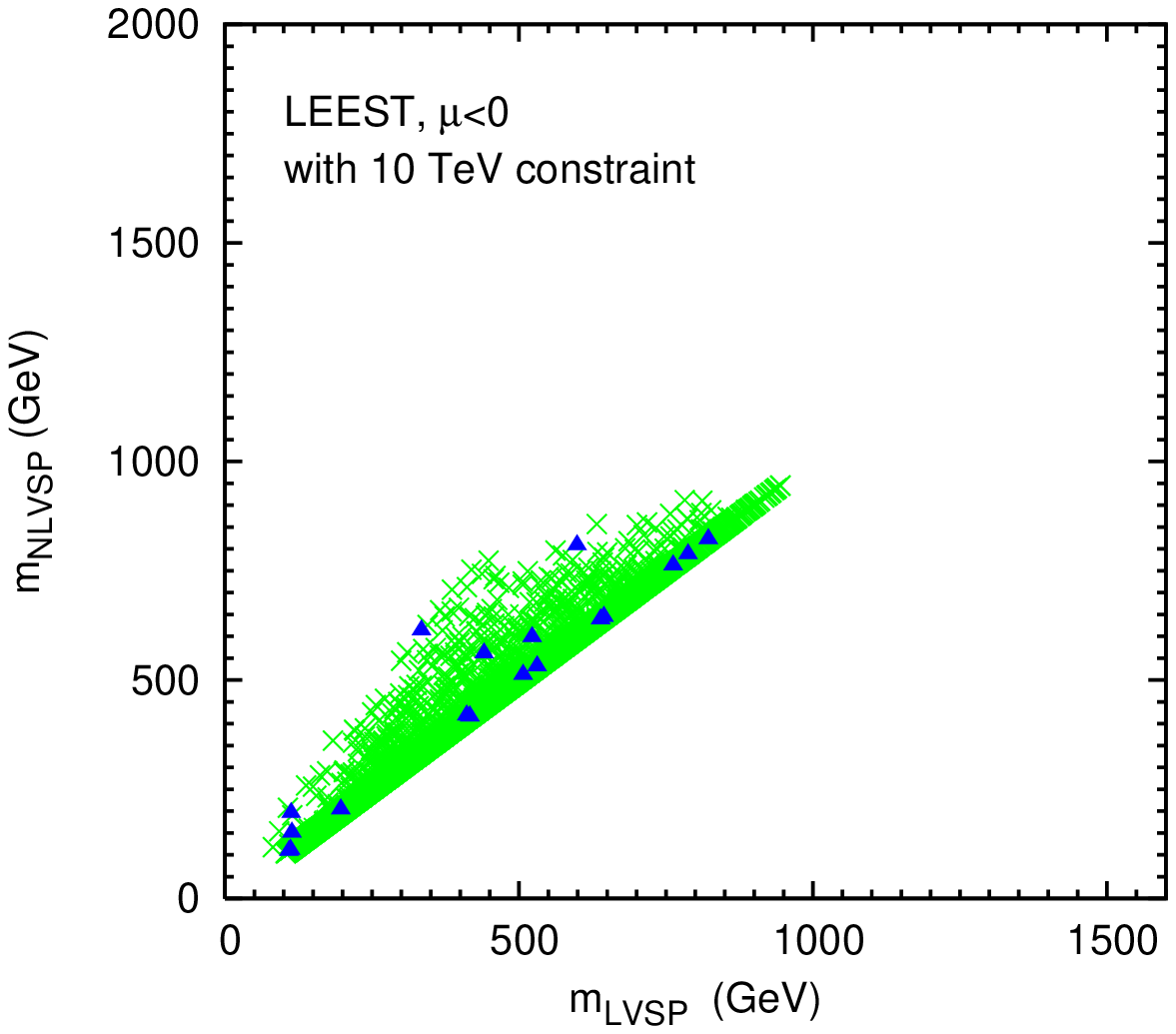,height=6.8cm}}
\end{center}
\caption{\it  
As in Fig.~\ref{fig:VSP10p}, but for $\mu < 0$.}
\label{fig:VSP10n}
\end{figure}

The panels of Fig.~\ref{fig:VSP10n} display the corresponding scatter
plots for the CMSSM, NUHM, LEEST and LEEST10 in the case that $\mu < 0$.
Although the scatter plots are qualitatively similar to those in
Fig.~\ref{fig:VSP10p}, there are some differences of detail between the
`sister' plots for the two signs of $\mu$. In particular, the upper bounds
on the LVSP and NLVSP masses are somewhat different: $(m_{LVSP},
m_{NLVSP}) \la (1350, 1400), (1400, 1400), (1600, 1600), (800, 800)$~GeV in
the (a) CMSSM, (b) NUHM, (c) LEEST and (d) LEEST10 cases, respectively.
In the CMSSM, the division between the dark (blue) triangles whose 
relic density is controlled by coannihilations and rapid s-channel annihilations
now occurs at a lower value of $m_{LVSP} \sim 500$ GeV. The two nearly 
linear tracks of points with large $m_{LVSP}$ corresponds to points with large
$m_0$ for which the LVSP and NLVSP are a nearly degenerate pair of charginos and 
neutralinos (lower track), and points with smaller $m_0$ where the LVSP, NLVSP pair is the
stau and selectron/smuon. 
However, the overall conclusions about the physics reaches of LCs with
different $E_{CM}$ are similar: low-energy LCs with $E_{CM} \le 1000$~GeV
reach part of the allowed parameter space, whereas a LC with $E_{CM} =
3200$~GeV would be `guaranteed' to find sparticles in all of these
models. The required centre-of-mass energies for each case are 
individually summarized in 
Table 1.

\begin{figure}
\begin{center}
\mbox{\epsfig{file=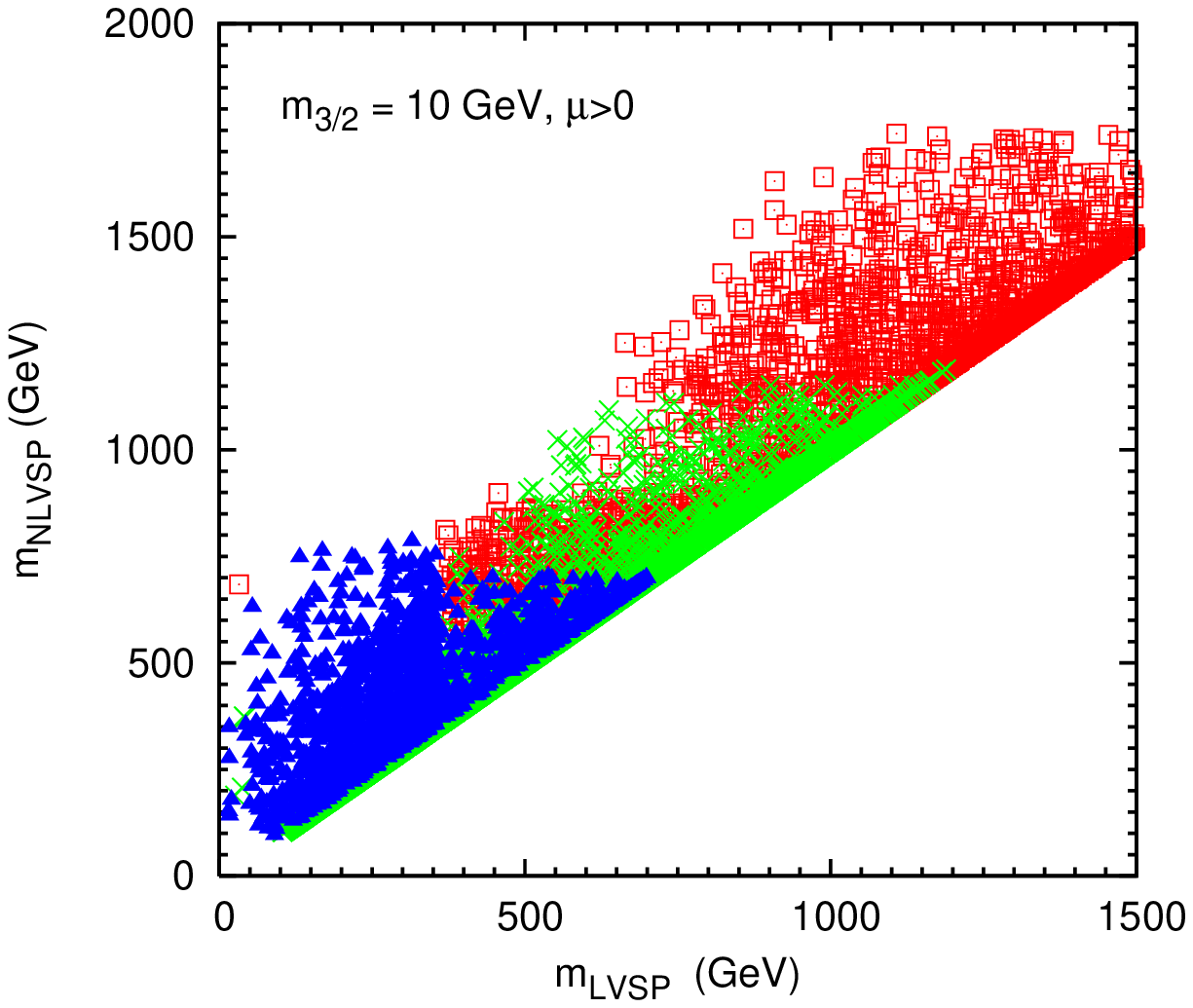,height=6.8cm}}
\mbox{\epsfig{file=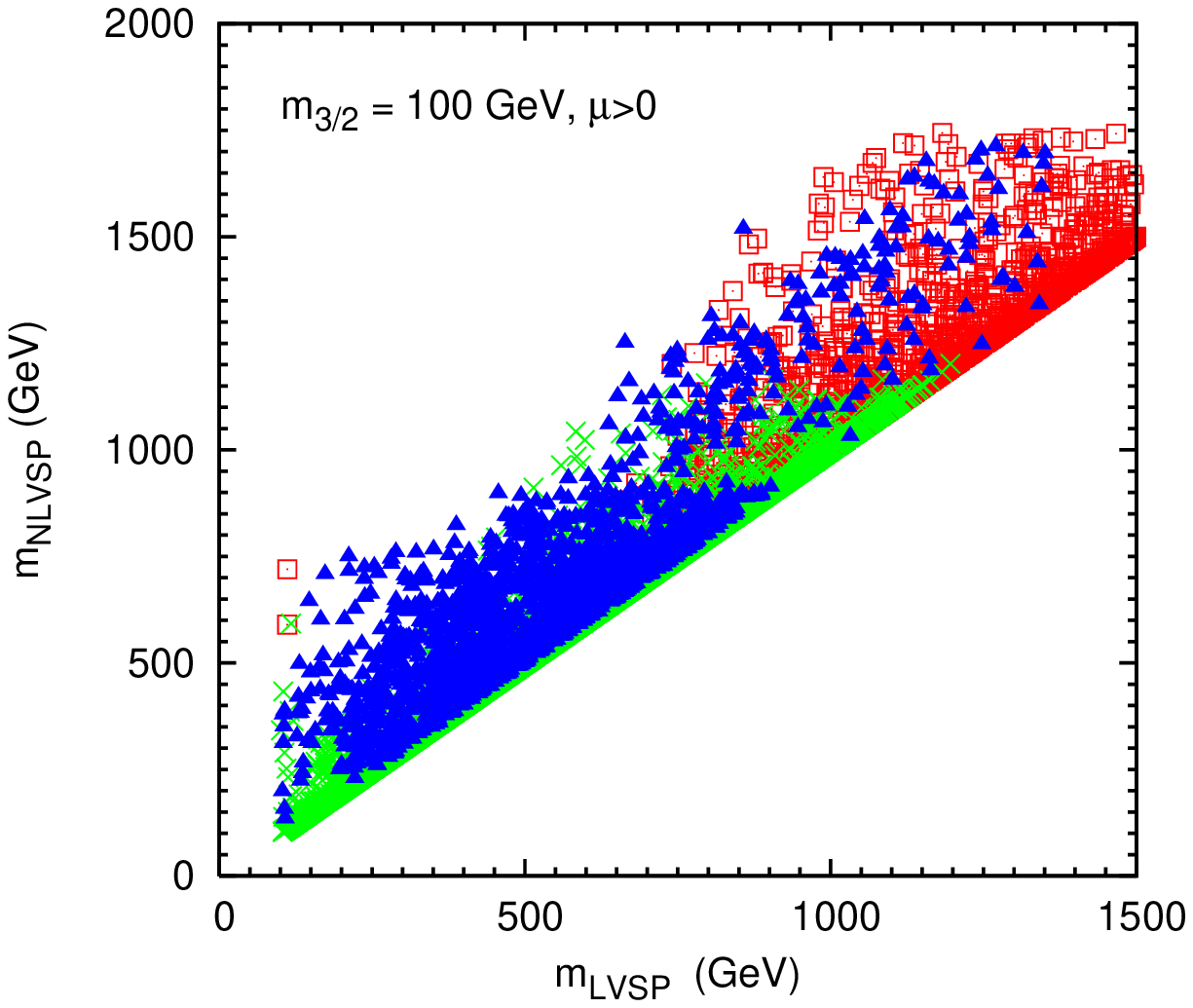,height=6.8cm}}
\end{center}
\begin{center}
\mbox{\epsfig{file=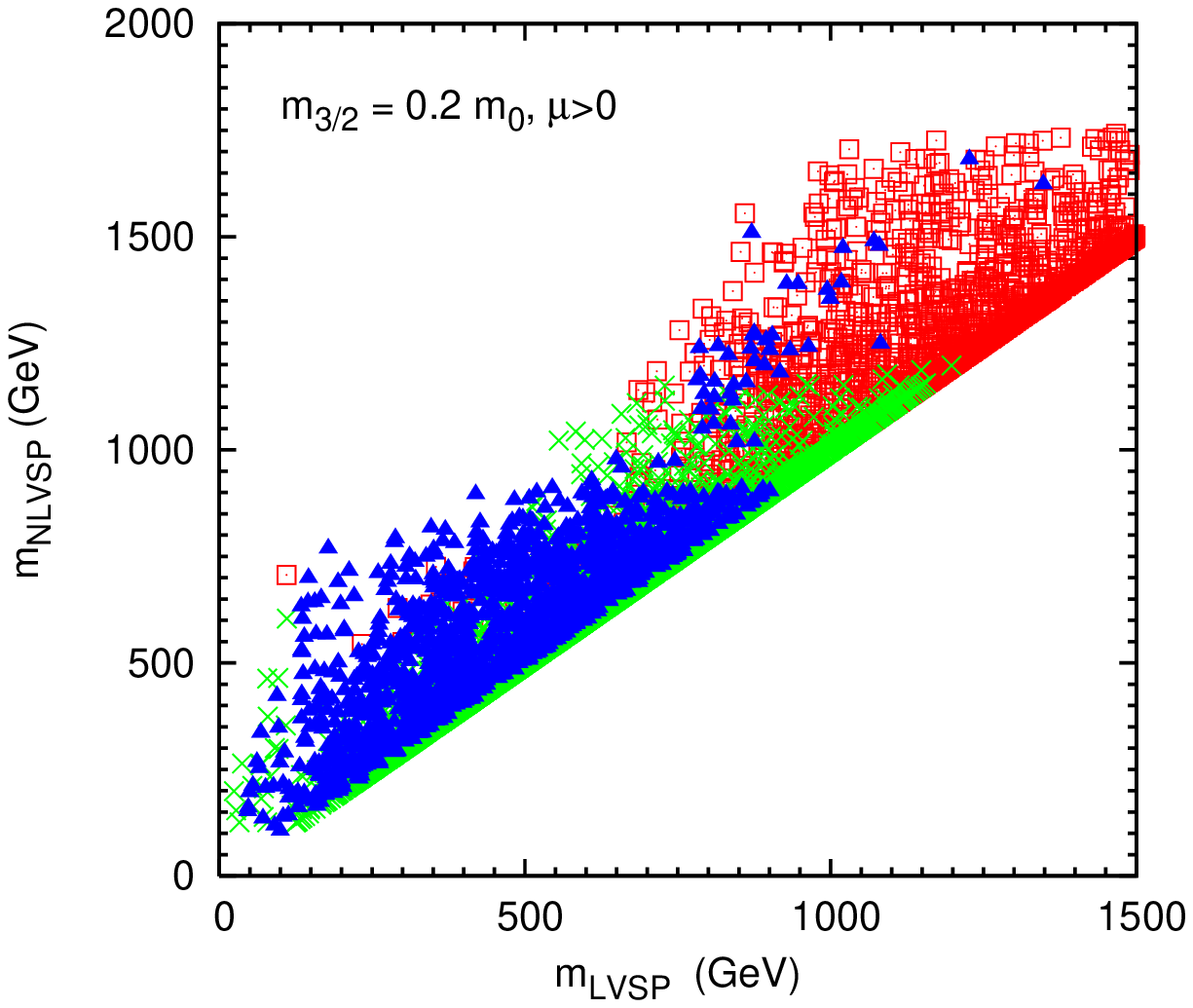,height=6.8cm}}
\mbox{\epsfig{file=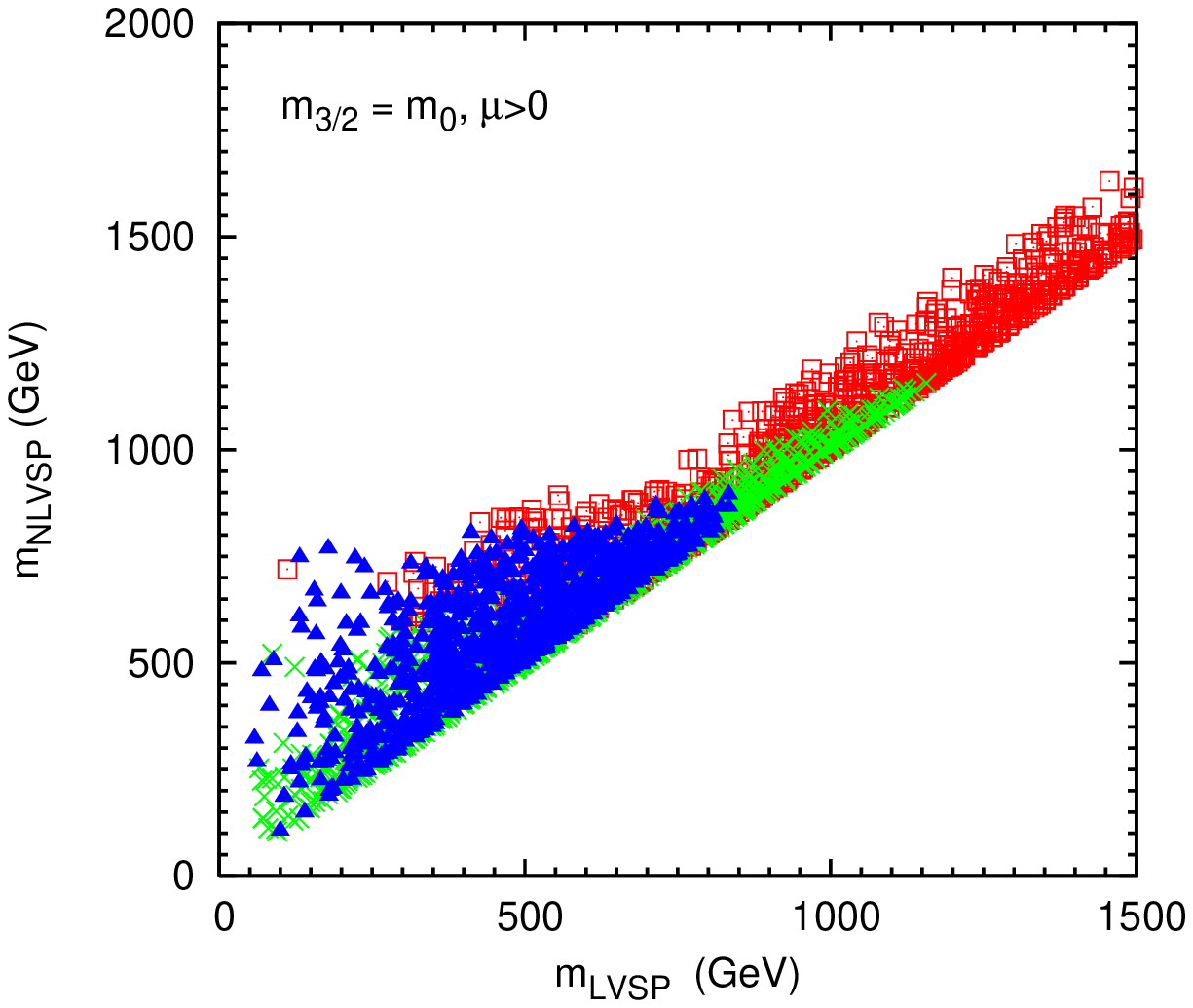,height=6.8cm}}
\end{center}
\caption{\it  
Scatter plots of the masses of the lightest visible
supersymmetric particle (LVSP) and the next-to-lightest visible
supersymmetric particle (NLVSP) in the GDM with (a) $m_{3/2} = 10$~GeV, 
(b) $m_{3/2} = 100$~GeV, (c) $m_{3/2} = 0.2 m_0$ and (d) $m_{3/2} = 
m_0$, all for $\mu > 0$. The darker (blue) 
triangles satisfy all the laboratory, astrophysical and cosmological 
constraints. For comparison, the dark (red) squares and medium-shaded 
(green) crosses respect the laboratory 
constraints, but not those imposed by astrophysics and cosmology. 
In addition, the (green) crosses represent models which are expected to be
visible at the LHC. } 
\label{fig:VSP10pGDM}
\end{figure}

The remaining figures display scatter plots in various scenarios with a
gravitino LSP, assuming scalar-mass universality. In the absence of any
better-tailored analysis, we use the same criterion~\cite{Baer} for
observability at the LHC.  We recall that the allowed regions of the
$(m_{1/2}, m_0) $ planes in such GDM scenarios are very different from
those allowed in the CMSSM~\cite{gdm}. Our own studies of the GDM have
been restricted to a few specific scenarios for the gravitino mass 
$m_{3/2}$~\cite{gdm}~\footnote{Moreover, the computer time required to 
generate a useful 
sample in the higher-dimensional space with $m_{3/2}$ a free parameter 
would be prohibitive.}. We only consider cases where the 
next-to-lightest supersymmetric
particle (NSP) has a lifetime exceeding $10^4$~s~\cite{CEFO}, since we
have not yet incorporated the effects of hadron showers in the early
Universe, which are expected to be important for shorter
lifetimes~\cite{hadr}. These limitations restrict our analysis here
artificially to portions of the GDM parameter space. For this reason, we
do not exclude the possibility that heavier LVSP and NLVSP masses might be
permitted, and the ranges of masses quoted below should be interpreted as
implying that a LC with $E_{CM}$ at least twice as large would be needed
for any `guarantee' of discovering supersymmetry in these scenarios. In 
the
specific case $m_{3/2} = 10$~GeV shown in panel (a) of
Fig.~\ref{fig:VSP10pGDM}, we find LVSP and NLVSP masses up to 700~GeV and
800~GeV respectively, implying that a LC with $E_{CM} \gappeq 1700$~GeV
would be needed for even a limited `guarantee' of discovery. However, this
case in particular suffers from our restriction on the NSP lifetime.  For
a fixed value of $m_0$, the $\tilde \tau_1$ mass is limited by the gaugino
mass, $m_{1/2}$, which is in turn limited by our restriction on the NSP
lifetime.  This causes most of the allowed points to appear below
$m_{LVSP} \la 400$ GeV, as occurs when the (LVSP, NLVSP) pair are either
($\tilde \tau_1, \chi$) or ($\tilde \tau_1, \tilde e_R$). However, some
extension beyond $m_{LVSP} \sim 400$ GeV is possible for larger values of
$m_0$. In these cases, the maximum allowed mass is determined by the
gravitino relic density constraint: $\Omega_{3/2} h^2 = (m_{3/2}/m_\chi)
\Omega_\chi h^2 < 0.129$, and the (LVSP, NLVSP) pair are either
($\chi^\pm_1, \chi_2$) or ($\tilde \tau_1, \tilde e$).

\begin{figure}
\begin{center}
\mbox{\epsfig{file=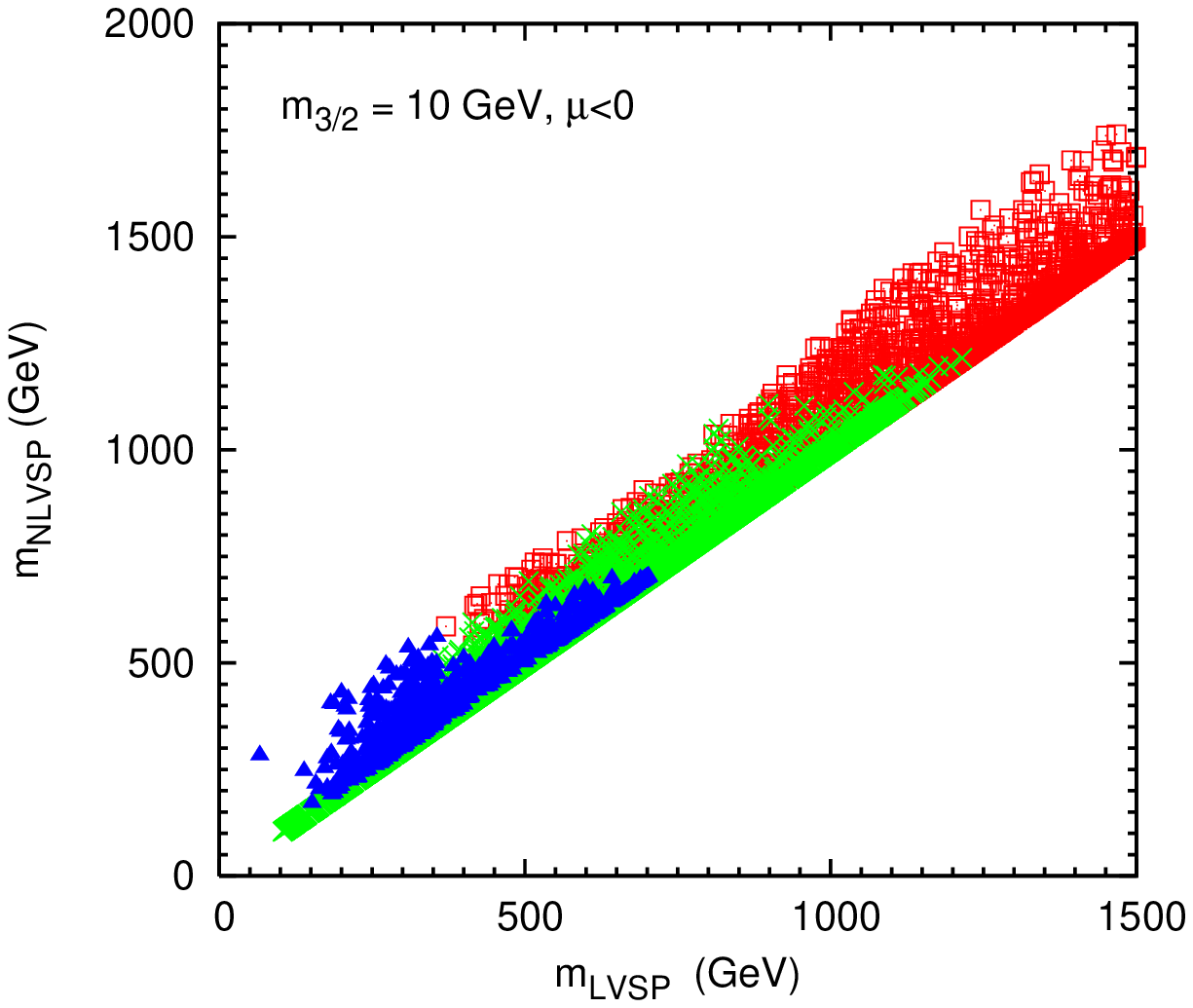,height=6.8cm}}
\mbox{\epsfig{file=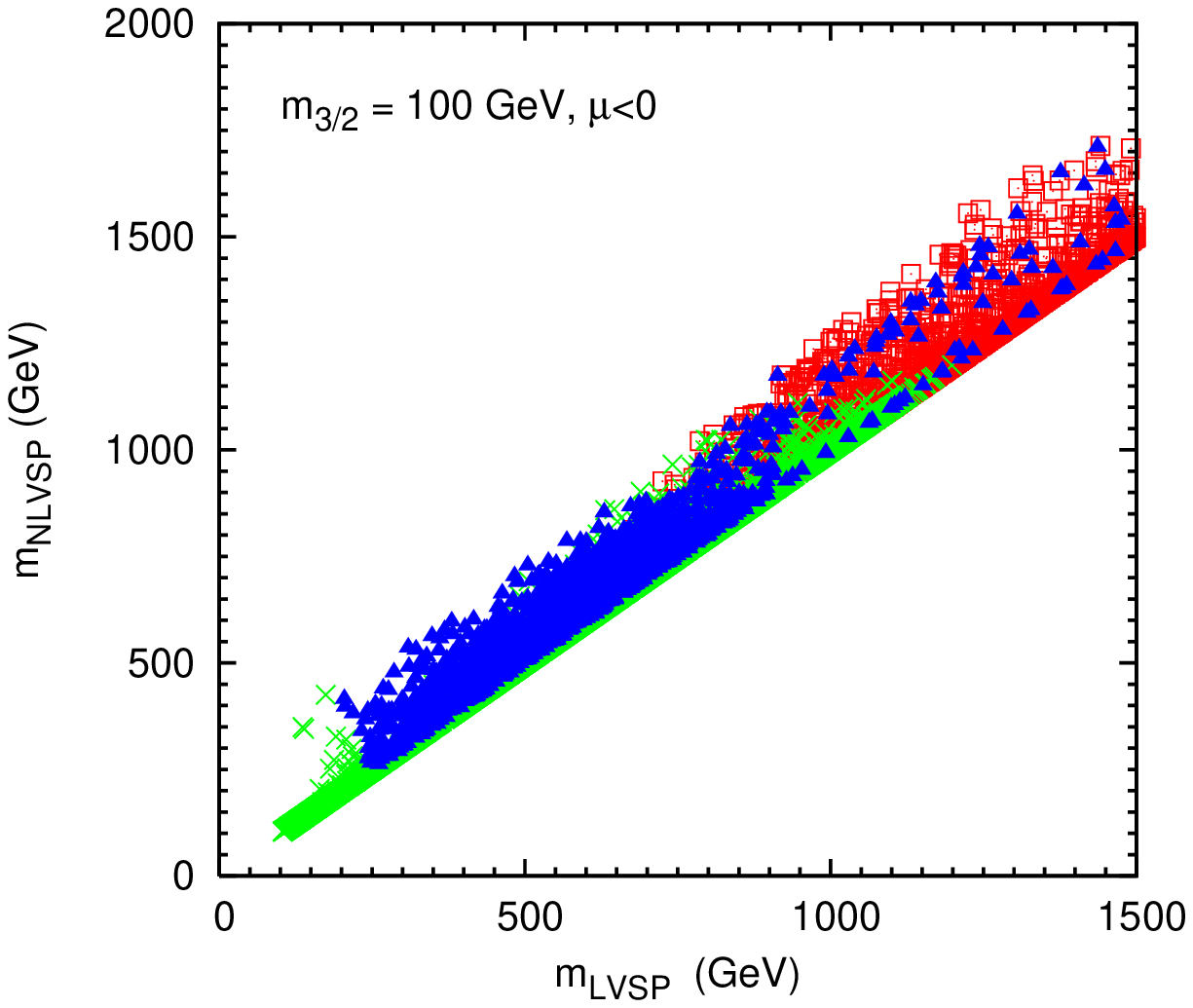,height=6.8cm}}
\end{center}
\begin{center}
\mbox{\epsfig{file=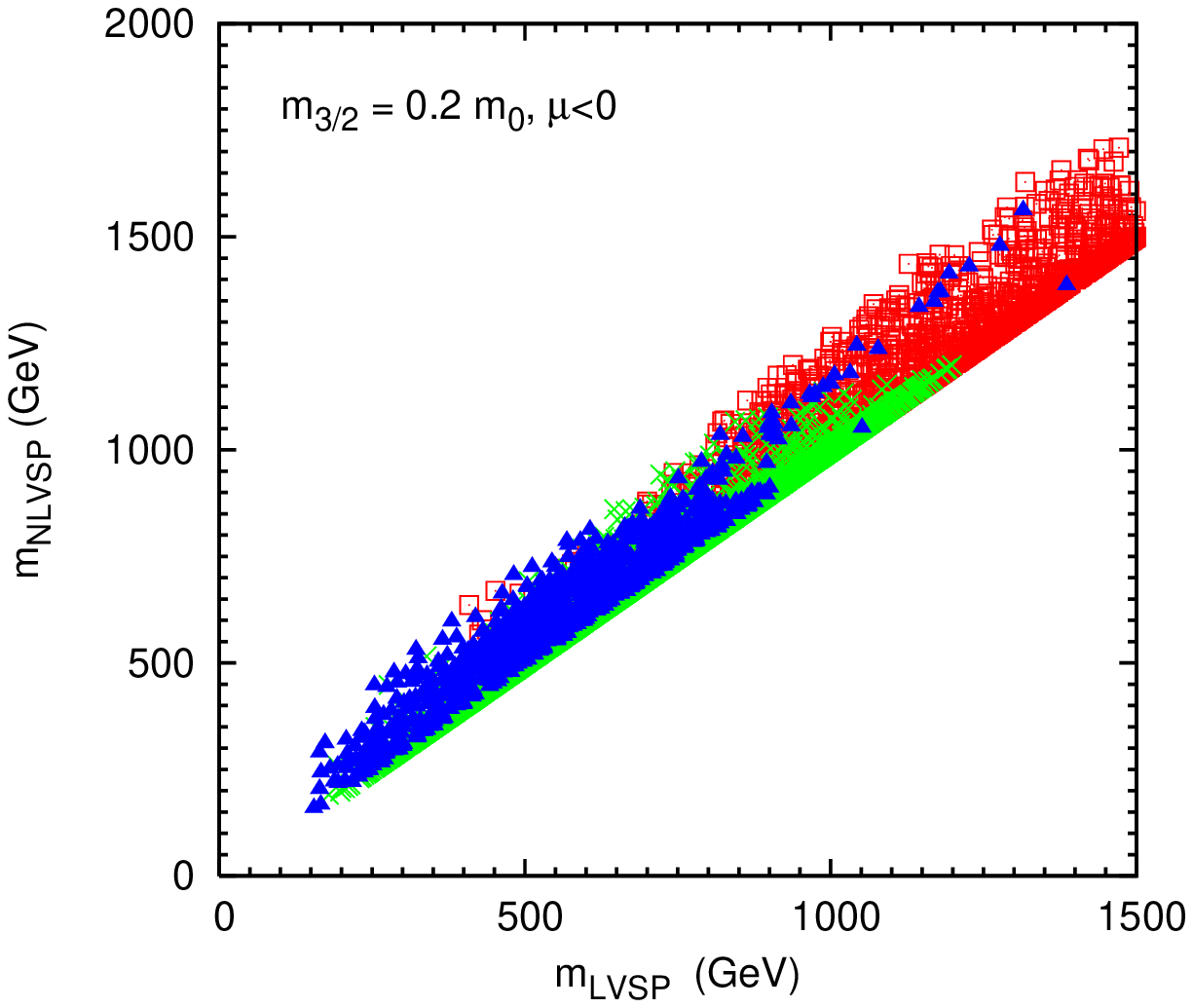,height=6.8cm}}
\mbox{\epsfig{file=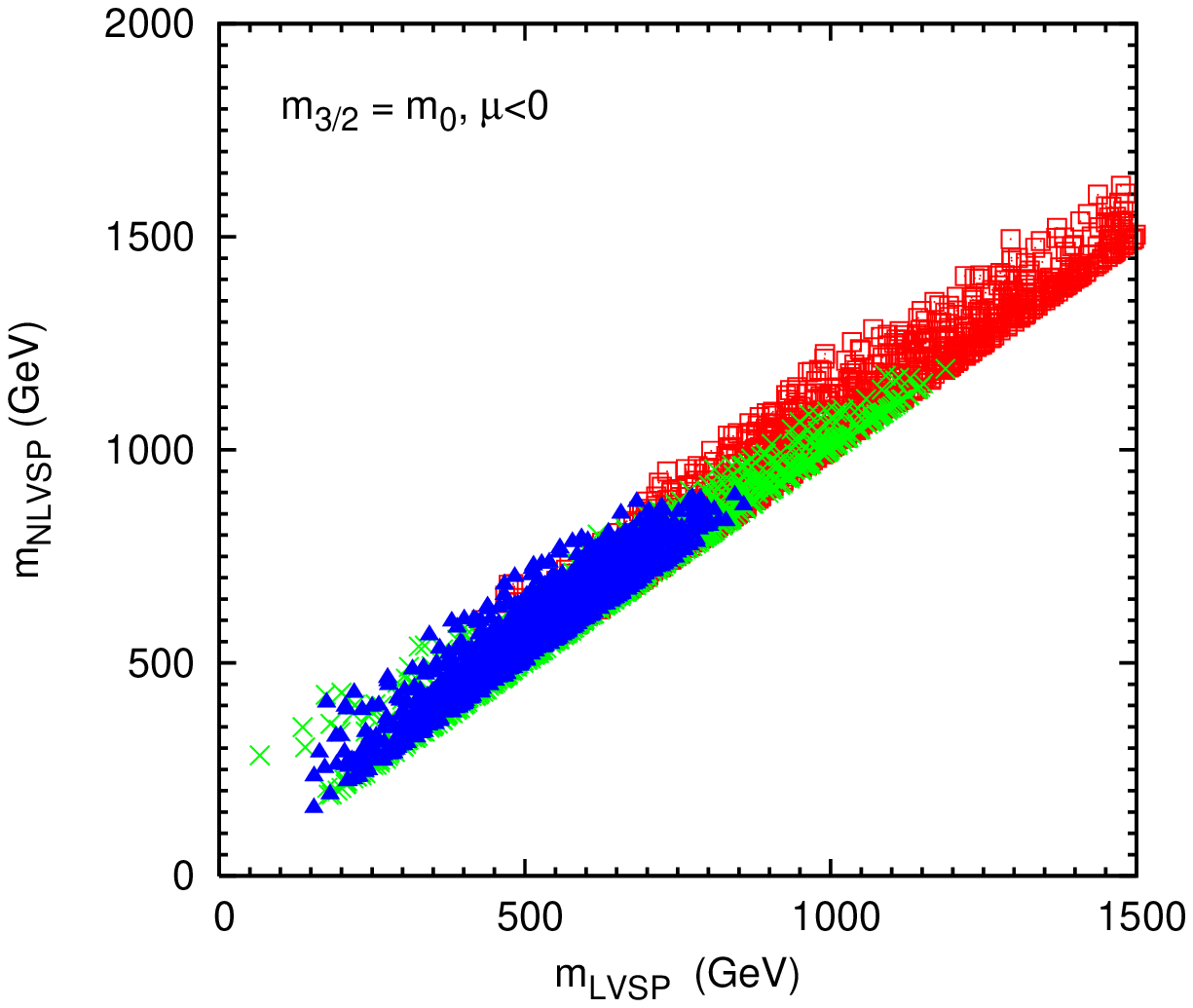,height=6.8cm}}
\end{center}
\caption{\it  
As in Fig.~\ref{fig:VSP10pGDM}, but for $\mu < 0$.}
\label{fig:VSP10nGDM}
\end{figure}

When $m_{3/2} = 100$ GeV, as shown in Fig.~\ref{fig:VSP10pGDM}(b), the
restriction due to the NSP lifetime is much less severe, and the LVSP and
NLVSP masses are allowed to roam to much higher values. Here, the
discontinuity at $m_{NLVSP} \sim 900$ GeV is simply a result of our chosen
range of $m_{1/2} < 2$ TeV. Although the dark (red) squares extend to
much higher masses, they have $m_{\chi} < m_{{\tilde \tau_1},{\tilde e_R}}$
and, for the most part, have $\Omega_{3/2} h^2$ above the WMAP limit.
However, a smattering of points with high $\tan \beta$ are allowed in the
rapid-annihilation funnel regions. Most of these points would not
be observable at the LHC. These same features are seen for
$m_{3/2} = 0.2 m_0$ in Fig.~\ref{fig:VSP10pGDM}(c) and in the
corresponding plots for $\mu < 0$ (Figs.~\ref{fig:VSP10nGDM}(b,c)).  
However, this feature is not found in panels (d) for either sign of $\mu$
(see, e.g.,~\cite{gdm}), as the funnel is no longer present when $m_{3/2}
= m_0$ because of the assumption that the LSP is the gravitino and the
limit on the on gravitino relic density.

We find no suggestion that a low-energy LC would be a safer bet in this
and other GDM scenarios than in the neutralino LSP scenarios discussed
earlier. In the cases $m_{3/2} = 100$~GeV, $m_{3/2} = 0.2 m_0$ and
$m_{3/2} = m_0$ shown in panels (b,c,d), respectively, we find $(m_{LVSP},
m_{NLVSP}) \la (1400, 1750), (1400, 1700), (850, 900)$ GeV, respectively. We
recall that minimal supergravity (mSUGRA) models have scalar-mass
universality, $m_{3/2} = m_0$ and a specific value for the universal
trilinear supersymmetry-breaking parameter $A$, and typically have
neutralino and gravitino LSPs in different regions of parameter
space~\cite{VCMSSM}. They are not equivalent to either the CMSSM or the
GDM scenario discussed here. This remark serves to emphasize that many
other scenarios for the masses of the MSSM particles and the gravitino
could be entertained, beyond those presented here, including also
scenarios with scalar masses that are non-universal to some degree, as
discussed earlier in connection with a neutralino LSP.  

The ranges of visible sparticle masses in the corresponding scenarios with
$\mu < 0$ are shown in Fig.~\ref{fig:VSP10nGDM}. Here we find in the cases
(a) $m_{3/2} = 10$~GeV, (b) $m_{3/2} = 100$~GeV, (c) $m_{3/2} = 0.2 m_0$
and (d) $m_{3/2} = m_0$, that $(m_{LVSP}, m_{NLVSP}) \la (700, 700),
(1500, 1700), (1400, 1600), (900, 900)$ GeV, respectively. The
centre-of-mass energies in each of these cases, as well as those for $\mu
> 0$, are summarized in Table 1.

\section{Prospects for Direct Detection of Supersymmetric Dark Matter}

One of the principal competitors with colliders for the discovery of 
supersymmetry is the search for astrophysical dark matter, assuming this 
to be composed of LSPs. Gravitino dark matter is very difficult to 
observe, but there are interesting prospects for detecting neutralino dark 
matter, either directly via scattering on nuclei, or indirectly via the 
products of annihilations in various astrophysical environments, such as 
the centres of the Earth, Sun or Galaxy, or in our galactic halo: for a 
recent review, see~\cite{joe}. Here, so 
as to minimize the astrophysical uncertainties, we focus on direct 
detection\footnote{Direct detection alone can not unambiguously
discover supersymmetry, as other non-supersymmetric dark matter candidates 
are possible.}.

There are two important contributions to generic $\chi$-nucleus
scattering, one that is spin-independent and related to quark 
contributions to
the nucleon mass, and one that is spin-dependent and related to quark
contributions to the nucleon spin. Since the former appears more promising 
in many experiments, we concentrate here on this spin-independent 
contribution.

Matrix elements for spin-independent $\chi$-nucleon scattering depend on
$<p| {\bar s} s |p>$, which may be estimated on the basis of the
$\sigma$ term in $\pi$-nucleon scattering. Recent evaluations of this
quantity appear to favour larger values than often assumed
previously~\cite{newsigma}, which may also be favoured by estimates based
on the possible spectroscopy of exotic baryons as treated in the chiral
soliton model~\cite{EKP}.  Accordingly, in this paper we use a larger
estimate of $<p| {\bar s} s |p>$ than in our previous work~\footnote{This
point is discussed in more detail in~\cite{EOSSsigma}.}: $y \equiv 2 <p|
{\bar s} s |p>/(<p| {\bar u} u |p>+ <p| {\bar d} d |p>) = 0.44$,
corresponding to $\sigma_{\pi N} = 64$ MeV.

Within the near future, searches for spin-independent $\chi$-nucleus
scattering are expected to reach a sensitivity $\sim 10^{-8}$~pb for a
range of $m_\chi$. We indicate in Figs.~\ref{fig:VSP10p} and
\ref{fig:VSP10n} by light (yellow) the randomly-selected models which have
cross sections above $10^{-8}$~pb. These populate the regions of low
$m_{LVSP}$ and $m_{NLVSP}$ that would be particularly accessible to a
low-energy LC. Note that in the CMSSM, the elastic scattering 
cross section for $\mu < 0$ is generally smaller than the corresponding
case when $\mu > 0$ (see e.g., \cite{eflo}). 
Furthermore, for $\mu < 0$, the $b \to s \gamma$ constraint
also eliminates points with large elastic scattering cross sections.
As such, no points in Fig. 
\ref{fig:VSP10n}a, rise above the $10^{-8}$ pb threshold. 

However, many of these models make an excessive contribution to $g_\mu -
2$. In fact if we applied the upper limit to $\delta a_\mu < 31 \times
10^{-10}$, roughly half of the light (yellow) circles are removed in panel
(a) of Fig.~\ref{fig:VSP10p} for the case of the CMSSM.  Of those
remaining, roughly half have a relic density below 0.0945. Not all the
supersymmetric models accessible to a low-energy LC would be detectable at
this cross-section level, so such a LC would certainly add value in this
region of parameter space, and the absence of a signal in this generation
of direct searches for supersymmetric dark matter should not be taken as
evidence that such a low-energy LC could not see supersymmetry. In the
NUHM (panel b) of Fig.~\ref{fig:VSP10p}, only a few ($\sim 5 \%$) of the
light (yellow) circles would be removed by the $g_\mu-2$ constraint.
However, in this case, most of the points ($\sim 80 \%$) have $\Omega h^2
< 0.0945$. These points typically correspond to a LSP which is
Higgsino-like. As a consequence the relic density is small, due to the
relatively large annihilation cross-section and the Higgs exchange channel
makes a strong contribution to the total elastic cross-section.

\section{Summary}

We have explored the prospects for discovering one or more supersymmetric
particles in a number of models with either a neutralino or a gravitino
LSP. We have considered various hypotheses for relations between soft
supersymmetry-breaking scalar masses with differing degrees of
universality. In all the models studied, we find that a low-energy LC with
$E_{CM} \le 1000$~GeV has a chance to produce and detect one or more
sparticles, but this cannot be guaranteed. However, a high-energy LC with
$E_{CM} \ge 3000$~GeV would be needed to `guarantee' the detection of
supersymmetry in neutralino LSP models, and we cannot exclude the
possibility that an even higher $E_{CM}$ might be required in some models
with a gravitino LSP.

It is clear that the naturalness of the electroweak symmetry-breaking
scale favours lower sparticle masses to some extent~\cite{EENZ}, but there 
is no clear
criterion how this aesthetic requirement should be imposed. One might
strike lucky with some search for supersymmetric dark matter, either
direct (as discussed here) or indirect, but this is not guaranteed, even
if the supersymmetry breaking scale is relatively low. The next clear
information on the sparticle mass scale may have to wait for data from the
LHC.

\vspace*{1cm}
\noindent{ {\bf Acknowledgments} } \\
\noindent 
The work of K.A.O., Y.S., and V.C.S. was supported in part
by DOE grant DE--FG02--94ER--40823.


\begin{thebibliography}{99}

\bibitem{EHNOS}
J. Ellis, J.S. Hagelin, D.V. Nanopoulos, K.A. Olive
and M. Srednicki, Nucl. Phys. B {\bf 238} (1984) 453; see also
H. Goldberg, Phys. Rev. Lett. {\bf 50} (1983) 1419.


\bibitem{ekn}
J.~R.~Ellis, J.~E.~Kim and D.~V.~Nanopoulos,
Phys.\ Lett.\ B {\bf 145} (1984) 181;
T.~Moroi, H.~Murayama and M.~Yamaguchi,
Phys.\ Lett.\ B {\bf 303} (1993) 289;
J.~R.~Ellis, D.~V.~Nanopoulos, K.~A.~Olive and S.~J.~Rey,
Astropart.\ Phys.\  {\bf 4} (1996) 371
[arXiv:hep-ph/9505438];
M.~Bolz, W.~Buchmuller and M.~Plumacher,
Phys.\ Lett.\ B {\bf 443} (1998) 209
[arXiv:hep-ph/9809381];
T.~Gherghetta, G.~F.~Giudice and A.~Riotto,
Phys.\ Lett.\ B {\bf 446} (1999) 28
[arXiv:hep-ph/9808401];
T.~Asaka, K.~Hamaguchi and K.~Suzuki,
Phys.\ Lett.\ B {\bf 490} (2000) 136
[arXiv:hep-ph/0005136];
M.~Fujii and T.~Yanagida,
Phys.\ Rev.\ D {\bf 66} (2002) 123515
[arXiv:hep-ph/0207339];
Phys.\ Lett.\ B {\bf 549} (2002) 273
[arXiv:hep-ph/0208191].

\bibitem{BBB}
M.~Bolz, A.~Brandenburg and W.~Buchmuller,
Nucl.\ Phys.\ B {\bf 606} (2001) 518
[arXiv:hep-ph/0012052];
W.~Buchmuller, K.~Hamaguchi and M.~Ratz,
Phys.\ Lett.\ B {\bf 574} (2003) 156
[arXiv:hep-ph/0307181].

\bibitem{gdm}
J.~R.~Ellis, K.~A.~Olive, Y.~Santoso and V.~C.~Spanos,
Phys.\ Lett.\ B {\bf 588} (2004) 7
[arXiv:hep-ph/0312262].


\bibitem{feng}
J.~L.~Feng, S.~Su and F.~Takayama,
arXiv:hep-ph/0404231;
arXiv:hep-ph/0404198;
J.~L.~Feng, A.~Rajaraman and F.~Takayama,
Phys.\ Rev.\ Lett.\  {\bf 91} (2003) 011302
[arXiv:hep-ph/0302215].

\bibitem{CEFO}
R.~H.~Cyburt, J.~R.~Ellis, B.~D.~Fields and K.~A.~Olive,
Phys.\ Rev.\ D {\bf 67} (2003) 103521
[arXiv:astro-ph/0211258].

\bibitem{us}
J.~R.~Ellis, K.~A.~Olive and Y.~Santoso,
New Jour.\ Phys.\  {\bf 4} (2002) 32
[arXiv:hep-ph/0202110];
J.~R.~Ellis, K.~A.~Olive, Y.~Santoso and V.~C.~Spanos,
Phys.\ Lett.\ B {\bf 565} (2003) 176
[arXiv:hep-ph/0303043];
J.~R.~Ellis, K.~A.~Olive, Y.~Santoso and V.~C.~Spanos,
Phys.\ Rev.\ D {\bf 69} (2004) 095004
[arXiv:hep-ph/0310356].



\bibitem{them}
V.~D.~Barger and C.~Kao,
Phys.\ Lett.\ B {\bf 518} (2001) 117
[arXiv:hep-ph/0106189];
L.~Roszkowski, R.~Ruiz de Austri and T.~Nihei,
JHEP {\bf 0108} (2001) 024
[arXiv:hep-ph/0106334];
A.~B.~Lahanas and V.~C.~Spanos,
Eur.\ Phys.\ J.\ C {\bf 23} (2002) 185
[arXiv:hep-ph/0106345];
U.~Chattopadhyay, A.~Corsetti and P.~Nath,
Phys.\ Rev.\ D {\bf 66} (2002) 035003
[arXiv:hep-ph/0201001];
H.~Baer, C.~Balazs, A.~Belyaev, J.~K.~Mizukoshi, X.~Tata and Y.~Wang,
JHEP {\bf 0207} (2002) 050
[arXiv:hep-ph/0205325];
R.~Arnowitt and B.~Dutta,
arXiv:hep-ph/0211417.
H.~Baer and C.~Balazs,
JCAP {\bf 0305} (2003) 006
[arXiv:hep-ph/0303114];
A.~B.~Lahanas and D.~V.~Nanopoulos,
Phys.\ Lett.\ B {\bf 568} (2003) 55
[arXiv:hep-ph/0303130];
U.~Chattopadhyay, A.~Corsetti and P.~Nath,
Phys.\ Rev.\ D {\bf 68} (2003) 035005
[arXiv:hep-ph/0303201];
C.~Munoz,
arXiv:hep-ph/0309346;
R.~Arnowitt, B.~Dutta and B.~Hu,
arXiv:hep-ph/0310103.

\bibitem{djou}
A.~Djouadi, M.~Drees and J.~L.~Kneur,
JHEP {\bf 0108} (2001) 055
[arXiv:hep-ph/0107316].

\bibitem{nonu}
M.~Drees, M.~M.~Nojiri, D.~P.~Roy and Y.~Yamada,
\PR D~{\bf 56} (1997) 276
[Erratum-ibid. D~{\bf 64} (1997) 039901]
[arXiv:hep-ph/9701219];
M.~Drees, Y.~G.~Kim, M.~M.~Nojiri, D.~Toya, K.~Hasuko and T.~Kobayashi,
\PR D~{\bf 63} (2001) 035008
[arXiv:hep-ph/0007202];
V.~Berezinsky, A.~Bottino, J.~R.~Ellis, N.~Fornengo, G.~Mignola and S.~Scopel,
{Astropart.\ Phys.}  {\bf 5} (1996) 1
[arXiv:hep-ph/9508249];
P.~Nath and R.~Arnowitt,
\PR D~{\bf 56} (1997) 2820
[arXiv:hep-ph/9701301];
A.~Bottino, F.~Donato, N.~Fornengo and S.~Scopel,
\PR D~{\bf 63} (2001) 125003
[arXiv:hep-ph/0010203]; 
S.~Profumo,
Phys.\ Rev.\ D {\bf 68} (2003) 015006
[arXiv:hep-ph/0304071].

\bibitem{nuhm}
J.~Ellis, K.~Olive and Y.~Santoso,
\PL B~{\bf 539} (2002) 107
[arXiv:hep-ph/0204192];
J.~R.~Ellis, T.~Falk, K.~A.~Olive and Y.~Santoso,
Nucl.\ Phys.\ B {\bf 652} (2003) 259
[arXiv:hep-ph/0210205].

\bibitem{LEEST}
J.~R.~Ellis, K.~A.~Olive, Y.~Santoso and V.~C.~Spanos,
Phys.\ Lett.\ B {\bf 573} (2003) 163
[arXiv:hep-ph/0308075].


\bibitem{related}
J.~R.~Ellis, G.~Ganis and K.~A.~Olive,
Phys.\ Lett.\ B {\bf 474} (2000) 314
[arXiv:hep-ph/9912324];
M.~Battaglia {\it et al.},
Eur.\ Phys.\ J.\ C {\bf 22} (2001) 535
[arXiv:hep-ph/0106204];
B.~C.~Allanach {\it et al.},
in {\it Proc. of the APS/DPF/DPB Summer Study on the Future of Particle Physics (Snowmass 2001) } ed. N.~Graf,
Eur.\ Phys.\ J.\ C {\bf 25} (2002) 113
[eConf {\bf C010630} (2001) P125]
[arXiv:hep-ph/0202233];
M.~Battaglia, A.~De Roeck, J.~R.~Ellis, F.~Gianotti, K.~A.~Olive and L.~Pape,
Eur.\ Phys.\ J.\ C {\bf 33} (2004) 273
[arXiv:hep-ph/0306219];
R.~Arnowitt, B.~Dutta, T.~Kamon and V.~Khotilovich,
arXiv:hep-ph/0308159;
H.~Baer, A.~Belyaev, T.~Krupovnickas and X.~Tata,
JHEP {\bf 0402} (2004) 007
[arXiv:hep-ph/0311351];
K.~Desch, J.~Kalinowski, G.~Moortgat-Pick, M.~M.~Nojiri and G.~Polesello,
JHEP {\bf 0402} (2004) 035
[arXiv:hep-ph/0312069];
W.~Buchmuller, K.~Hamaguchi, M.~Ratz and T.~Yanagida,
Phys.\ Lett.\ B {\bf 588} (2004) 90
[arXiv:hep-ph/0402179];
B.~C.~Allanach, G.~A.~Blair, S.~Kraml, H.~U.~Martyn, G.~Polesello, W.~Porod and P.~M.~Zerwas,
arXiv:hep-ph/0403133;
R.~Lafaye, T.~Plehn and D.~Zerwas,
arXiv:hep-ph/0404282;
H.~Baer, T.~Krupovnickas and X.~Tata,
JHEP {\bf 0406} (2004) 061
[arXiv:hep-ph/0405058].

\bibitem{EENZ}
J.~R.~Ellis, K.~Enqvist, D.~V.~Nanopoulos and F.~Zwirner,
Mod.\ Phys.\ Lett.\ A {\bf 1} (1986) 57;
R.~Barbieri and G.~F.~Giudice,
Nucl.\ Phys.\ B {\bf 306} (1988) 63.

\bibitem{mtop}
CDF Collaboration, D0 Collaboration and Tevatron Electroweak Working Group
arXiv:hep-ex/0404010.

\bibitem{FeynHiggs}
S.~Heinemeyer, W.~Hollik and G.~Weiglein,
Comput.\ Phys.\ Commun.\  {\bf 124} (2000) 76
[arXiv:hep-ph/9812320];
S.~Heinemeyer, W.~Hollik and G.~Weiglein,
Eur.\ Phys.\ J.\ C {\bf 9} (1999) 343
[arXiv:hep-ph/9812472].

\bibitem{funnels}
M.~Drees and M.~M.~Nojiri,
Phys.\ Rev.\ D {\bf 47} (1993) 376
[arXiv:hep-ph/9207234];
H.~Baer and M.~Brhlik,
Phys.\ Rev.\ D {\bf 53} (1996) 597
[arXiv:hep-ph/9508321];
A.~B.~Lahanas, D.~V.~Nanopoulos and V.~C.~Spanos,
Phys.\ Rev.\ D {\bf 62} (2000) 023515
[arXiv:hep-ph/9909497];
Mod.\ Phys.\ Lett.\ A {\bf 16} (2001) 1229
[arXiv:hep-ph/0009065];
H.~Baer, M.~Brhlik, M.~A.~Diaz, J.~Ferrandis, P.~Mercadante,  
P.~Quintana and X.~Tata,
Phys.\ Rev.\ D {\bf 63} (2001) 015007
[arXiv:hep-ph/0005027];
J.~R.~Ellis, T.~Falk, G.~Ganis, K.~A.~Olive and M.~Srednicki,
Phys.\ Lett.\ B {\bf 510} (2001) 236
[arXiv:hep-ph/0102098].

\bibitem{fp}
J.~L.~Feng, K.~T.~Matchev and T.~Moroi,
Phys.\ Rev.\ Lett.\  {\bf 84} (2000) 2322
[arXiv:hep-ph/9908309];
J.~L.~Feng, K.~T.~Matchev and T.~Moroi,
Phys.\ Rev.\ D {\bf 61} (2000) 075005
[arXiv:hep-ph/9909334];
J.~L.~Feng, K.~T.~Matchev and F.~Wilczek,
Phys.\ Lett.\ B {\bf 482} (2000) 388
[arXiv:hep-ph/0004043].

\bibitem{rs}
A.~Romanino and A.~Strumia,
Phys.\ Lett.\ B {\bf 487} (2000) 165
[arXiv:hep-ph/9912301].

\bibitem{g-2}
G.~W.~Bennett {\it et al.}  [Muon g-2 Collaboration],
Phys.\ Rev.\ Lett.\  {\bf 92} (2004) 161802
[arXiv:hep-ex/0401008].

\bibitem{wmap}
C.~L.~Bennett {\it et al.},
Astrophys.\ J.\ Suppl.\  {\bf 148} (2003) 1
[arXiv:astro-ph/0302207].

\bibitem{kmy}
M.~Kawasaki, T.~Moroi and T.~Yanagida,
Phys.\ Lett.\ B {\bf 370} (1996) 52
[arXiv:hep-ph/9509399].

\bibitem{Baer}
H.~Baer, C.~Balazs, A.~Belyaev, T.~Krupovnickas and X.~Tata,
JHEP {\bf 0306} (2003) 054
[arXiv:hep-ph/0304303].

\bibitem{stopco}
C.~Boehm, A.~Djouadi and M.~Drees,
Phys.\ Rev.\ D {\bf 62} (2000) 035012
[arXiv:hep-ph/9911496];
J.~R.~Ellis, K.~A.~Olive and Y.~Santoso,
Astropart.\ Phys.\  {\bf 18} (2003) 395
[arXiv:hep-ph/0112113].




\bibitem{ENNC}
J.~R.~Ellis and D.~V.~Nanopoulos,
Phys.\ Lett.\ B {\bf 110} (1982) 44;
R.~Barbieri and R.~Gatto,
Phys.\ Lett.\ B {\bf 110} (1982) 211.

\bibitem{hadr}
M.~H.~Reno and D.~Seckel,
Phys.\ Rev.\ D {\bf 37} (1988) 3441;
S.~Dimopoulos, R.~Esmailzadeh, L.~J.~Hall and G.~D.~Starkman,
Nucl.\ Phys.\ B {\bf 311} (1989) 699;
K.~Kohri,
Phys.\ Rev.\ D {\bf 64} (2001) 043515
[arXiv:astro-ph/0103411];
M.~Kawasaki, K.~Kohri and T.~Moroi,
arXiv:astro-ph/0402490.

\bibitem{VCMSSM}
J.~Ellis, K.~A.~Olive, Y.~Santoso and V.~C.~Spanos,
arXiv:hep-ph/0405110.

\bibitem{joe}
G.~Bertone, D.~Hooper and J.~Silk,
arXiv:hep-ph/0404175.

\bibitem{newsigma}
M.~M.~Pavan, I.~I.~Strakovsky, R.~L.~Workman and R.~A.~Arndt,
PiN Newslett.\  {\bf 16}, 110 (2002)
[arXiv:hep-ph/0111066];
P.~Schweitzer,
arXiv:hep-ph/0312376.
see also: A.~Bottino, F.~Donato, N.~Fornengo and S.~Scopel,
Astropart.\ Phys.\  {\bf 13}, 215 (2000)
[arXiv:hep-ph/9909228];
Astropart.\ Phys.\  {\bf 18}, 205 (2002)
[arXiv:hep-ph/0111229];
E.~Accomando, R.~Arnowitt, B.~Dutta and Y.~Santoso,
Nucl.\ Phys.\ B {\bf 585}, 124 (2000)
[arXiv:hep-ph/0001019];
R.~Arnowitt, B.~Dutta and Y.~Santoso,
arXiv:hep-ph/0005154.



\bibitem{EKP}
J.~R.~Ellis, M.~Karliner and M.~Praszalowicz,
JHEP {\bf 0405} (2004) 002
[arXiv:hep-ph/0401127].

\bibitem{EOSSsigma}
J.~R.~Ellis, K.~A.~Olive, Y.~Santoso and V.~C.~Spanos,
in preparation.

\bibitem{eflo}
M.~Drees and M.~Nojiri, Phys.\ Rev.\ {\bf D48} (1993) 3483;
J.~R.~Ellis, A.~Ferstl and K.~A.~Olive,
Phys.\ Lett.\ B {\bf 481}, 304 (2000)
[arXiv:hep-ph/0001005];
J.~R.~Ellis, A.~Ferstl and K.~A.~Olive,
Phys.\ Rev.\ D {\bf 63}, 065016 (2001)
[arXiv:hep-ph/0007113].











\end{thebibliography}
\end{document}